\documentclass[a4paper]{article}
\pdfoutput=1
\usepackage[utf8]{inputenc}
\usepackage{a4wide} 
\usepackage{authblk}
\usepackage{url}
\usepackage[colorlinks=true,linkcolor=blue,citecolor=blue,urlcolor=blue]{hyperref}
\usepackage[pdftex]{graphicx,color}	
\usepackage{multirow}
\usepackage{soul}
\usepackage{amsmath}
  
\newcommand{\beginsupplement}{%
        \setcounter{table}{0}
        \renewcommand{\thetable}{S\arabic{table}}%
        \setcounter{figure}{0}
        \renewcommand{\thefigure}{S\arabic{figure}}%
     }  

\begin{document}

\title{Modelling plausible scenarios for the Omicron SARS-CoV-2 variant from early-stage surveillance}
\author[1]{Christopher J.\ Banks}
\author[1]{Ewan Colman}
\author[1]{Anthony J.\ Wood}
\author[2]{Thomas Doherty}
\author[1,3,*]{Rowland R.\ Kao}
\affil[1]{\small Roslin Institute, University of Edinburgh}
\affil[2]{\small Department of Mathematics and Statistics, University of Strathclyde}
\affil[3]{\small Royal (Dick) School of Veterinary Studies, University of Edinburgh}
\affil[*]{\small\texttt Correspondence: rowland.kao@ed.ac.uk}

\maketitle

\begin{abstract}
We used a spatially explicit agent-based model of SARS-CoV-2 transmission combined with spatially fine-grained COVID-19 observation data from Public Health Scotland to investigate the initial rise of the Omicron (BA.1) variant of concern. We evaluated plausible scenarios for transmission rate advantage and vaccine immune escape relative to the Delta variant based on the data that would have been available at that time. We also explored possible outcomes of different levels of imposed non-pharmaceutical intervention. The initial results of these scenarios were used to inform the Scottish Government in the early outbreak stages of the Omicron variant.

Using the model with parameters fit over the Delta variant epidemic, some initial assumptions about Omicron transmission rate advantage and vaccine escape, and a simple growth rate fitting procedure, we were able to capture the initial outbreak dynamics for Omicron. We found that the modelled dynamics hold up to retrospective scrutiny. The modelled imposition of extra non-pharmaceutical interventions planned by the Scottish Government at the time would likely have little effect in light of the transmission rate advantage held by the Omicron variant and the fact that the planned interventions would have occurred too late in the outbreak's trajectory. Finally, we found that any  assumptions made about the projected distribution of vaccines in the model population had little bearing on the outcome, in terms of outbreak size and timing.  Instead, it was the landscape of prior immunity that was most important. 
\end{abstract}

\section{Introduction}
The B.1.1.529 SARS-CoV-2 variant was first detected in South Africa and reported to the World Health Organisation on 24th November 2021; it was designated the Omicron variant of concern (VOC) or ``Omicron'' two days later~\cite{worldhealthorganisation2021}. The first cases of Omicron were detected in Scotland around 29th November~\cite{scottishgovernment2021}. Globally Omicron was associated with rapid spread and increase in case numbers, most likely due to some combination of increased transmission rate potential and increased vaccine escape~\cite{pulliam2022}. However greater infectiousness was somewhat offset by a reduction in outcome severity.

Upon the introduction of the variant to Scotland there was an urgent need to understand how the dynamics of transmission could put extra pressure on the already stressed National Health Service. It was also necessary to estimate the effect of any potential control measures to reduce the impact of a rapid outbreak of a new VOC. The new variant was particularly concerning, owing to multiple new mutations in the spike protein of the virus, and concern was raised about the potential for an increase in vaccine escape and more effective transmission. Indications from the outbreak in South Africa~\cite{pulliam2022} were confounded by low overall vaccine uptake in that country, therefore assumptions made on that basis had to be considered weak.

To address this, we adapted an existing simulation model of SARS-CoV-2 transmission in Scotland, SCoVMod~\cite{banks2022}, to investigate the rise of the Omicron VOC in the presence of previous variants, in order to evaluate plausible scenarios for transmission rate advantage and vaccine immune escape relative to the Delta VOC (``Delta''). We explored possible outcomes of different levels of imposed non-pharmaceutical interventions (NPI's) and booster vaccination, in order to provide insight into the possible severity of the epidemic in terms of the probable number of infections. In order to avoid making assumptions about vaccine escape and transmission rate advantage, we took a range of scenarios and the considered only those that fit closely with the already observed outbreak characteristics.

The initial results of these scenarios were used to inform the Scottish Government and intended to aid policy decision making in the early outbreak stages of the variant~\cite{scottishgovernment2022b}.

This model we use is similar in scope to other UK-based early-response scenario models, for example the models developed by the London School of Hygiene and Tropical Medicine~\cite{Davies2020,Davies2021a,Davies2021,barnard2021}, the University of Warwick~\cite{Keeling2022b,keeling2021,Keeling2021a,Moore2021}, and Imperial College London~\cite{Knock2021,Sonabend2021}.
However, it is an individual-based simulation model and also includes explicit inter-location mobility patterns derived from national statistical datasets, with finer grained data resolution than any other published UK-based model. This allows us to include, for example, regional variation in past exposure and effects of demography on transmission patterns as a natural feature of the model. 

We make use of the high levels of data granularity made available to us by the Scottish government which allows for spatially heterogeneous patterns of transmission to be compared to the distribution of cases. Prior immunity (to Delta), the movement-related connections between areas (both in and out of lockdowns), and spatial heterogeneity in health equity, as indicated by the Scottish Index of Multiple Deprivation Health Index have all been found to have significance descriptors of case distribution both in prior studies~\cite{banks2022,Wood2023} and by the finding that transmission trees generated by the model show a significant effect of Health Index deprivation on transmission patterns. We make similar assumptions on vaccine efficacy and escape to other models existing at the time of the onset of Omicron, but our model inherently contains fine detail about local characteristics of vaccine uptake and prior immunity allowing greater confidence when considering possible levels of vaccine escape.

\section{Methods}
Using the SCoVMod simulation model~\cite{banks2022}, we fit model parameters to the explicit pattern of recorded cases across all local authorities in Scotland, considering the period from August to December 2021. This was a period in which Delta was dominant~\cite{sheikh2021} and over which the COVID-19 epidemic in Scotland can be considered as a single infection process. We introduced a variant infection (representing Omicron), and then modelled the period from the 11th December to the end of March 2022. While considerable evidence on the differences between Omicron and Delta soon became available, here we assume knowledge only available up to 11th December when this analysis was initiated, to demonstrate an approach that is relevant to early outbreak analysis. We examine a range of scenarios with either fixed parameters or multipliers of the Delta parameters for the variant process.

In order to fit the model to observed data our simulated epidemics are compared to the spatio-temporal pattern of COVID-19 incidence in Scotland. Non-observable parameters were estimated using the number of infections estimated in the population. Incidence of SARS-CoV-2, here defined as the number of new infections each day, was not directly obtainable from surveillance data and therefore needs to be estimated. The number of confirmed cases found through PCR and lateral flow device tests is a useful indicator of incidence, however, a large proportion of infections would not be discovered through testing~\cite{colman2023}.

At local scales it is not appropriate to apply the same ascertainment rate to all sub-regions. This is particularly true in a heterogeneous population like Scotland where infection levels, access to testing, and test seeking propensity vary greatly between local authority areas. Hence we also introduce a novel method to estimate the incidence in each sub-region of Scotland. We derive a formula that takes the number of positive and negative PCR tests across the nation as input, and gives an estimate of the number of people who would test positive for SARS-CoV-2 in each sub-region on any given day. We then re-scale the prevalence estimate to form our estimate of incidence, which is used as our observed incidence in the model-fitting process.

\subsection{Data}
Data for fitting Delta, estimating the seeding of Omicron, and the distribution of COVID-19 vaccines was supplied by Public Health Scotland's  eDRIS team~\cite{publichealthscotland2022}. We differentiate between Delta and Omicron using S-Gene Target Failure (SGTF) in PCR tests as a proxy for Omicron infections. Community tests where SGTF is measured represent approximately 80\% of all tests at the time of running these scenarios. We also used publicly available Scottish census data from National Records for Scotland (NRS)~\cite{scottishgovernment2022}. We used datazone (DZ) level resolution where DZs are population census units of approximately 500 to 1,000 residents. The data for assignment of individuals to work locations is drawn from the NRS Census Flows data~\cite{ukdataservice2022}, Table WU01UK, which provides origin/destination workplace data for the population from the 2011 census. We adjust these with respect to the 2018 population estimates.

Age demographics and movement to work patterns are available at the level of Census Output Areas (OA), each of which contains approximately 20 households or 50 people~\cite{nationalrecordsofscotland2022}. Census data on the Scottish Index of Multiple Deprivation (SIMD)~\cite{scottishgovernment2022a} considers multiple relative deprivation measures and combines them into a single value. Deprivation data are publicly available at the DZ level.

We also used publicly available data from Google to estimate mobility levels over time, with respect to commuting patterns~\cite{googleinc.2022}.

\subsection{Model}
SCoVMod is an explicitly spatial agent-based simulation model that accounts for recorded commuter patterns and additional local movements that are intended to capture non-work interactions such as recreation, shopping, and school. These movements are further modulated by the recorded time-varying mobility statistics, and geographically explicit population age structures. Whilst this does not capture all human movement, we assume that commuting patterns capture a large proportion of the long-range mobility and that local movements at least partially capture the non-commuting population travelling to shops, schools, and other local community movements. This is a similar assumption regarding mobility to the one used in other studies, including e.g.\ Barnard et al.~\cite{barnard2021}. The model also uses deprivation metrics to account for spatial heterogeneity in outcome likelihoods. The model is parameterised against an estimation of the number of COVID-19 infections that we describe below. The model parameters were inferred using the well-established Sequential Monte Carlo approach to Approximate Bayesian Computation (ABC-SMC)~\cite{Toni2009}. The outputs of a range of scenarios and their projections were then used to estimate a plausible range of hospital admissions from variant cases.

The model output is summarised at national and Council Area levels. Council Areas are the largest administrative units into which Scotland is divided of which there are 32.

The core of the simulation model breaks down into the following parts:
\begin{itemize}
\item Local transmission---a homogeneous mixing compartmental model for each OA of the country;
\item National transmission---a network-based simulation of the movement of individuals between OAs;
\item Parameter inference---a Bayesian estimation of the parameters for local transmission, this also involves a model of infection incidence used as the observed value for inference;
\item Transmission rate over time adjustment---the modulation of both local and national transmission to simulate non-pharmaceutical interventions and other changes in transmission rates over time.
\end{itemize}

The compartmental model considers key aspects of COVID-19 epidemiology including phases for latent infection, infectious and mildly infected (showing few or no clinical signs) and severely infected (with substantial clinical signs) individuals,
hospitalised, recovered and died, similar to other investigations~\cite{Arenas2020,DiDomenico2020}. These epidemiological processes are captured as individual disease states (Figure~\ref{fig:model-structure}). Individuals are also stratified into three age groups: young (0--15), adult (16--64) and elderly (65+). Within-OA transmission is assumed to be homogeneously mixed while between-OA transmission is determined by the empirical age-specific patterns of home and work contact (creating day/night patterns of contact). We do not consider overnight shifts in location or introductions from outside Scotland beyond the impact on the initial seeding. 

Deprivation is also known to influence COVID-19 transmission. We therefore adjust transmission rates in the model according to the average SIMD health index in the local Council Area. 

Population mobility patterns are determined by the patterns of movements to work recorded in Scottish Census data. We assume that only adults contribute to commuter movement, in the daytime. The remaining proportion of adults and all young and elderly individuals are assumed to move primarily within their local OAs, but also in some proportion to nearby OAs, to account for non-work movements (e.g.\ school, shopping, recreation). Finally, movement in the model is restricted to healthy and exposed or mildly symptomatic individuals; severely infected and hospitalised individuals do not move.

Individuals move within the spatial structure of the model on a day/night cycle. During the day those who commute are moved to their work location, and those others who move locally are moved to their other daytime location, transmission occurs within these locations and in the home locations for those who remain. During the night all individuals return to their home locations and transmission occurs within the home locations. The day/night pattern results in two transmission rates---the day rate where adults have moved to work locations and others have potentially moved to nearby locations, and the night rate where all individuals have returned to their home locations.

\subsubsection*{Local transmission (within-OA)}
\begin{figure}
  \centering \includegraphics[width=0.99\textwidth]{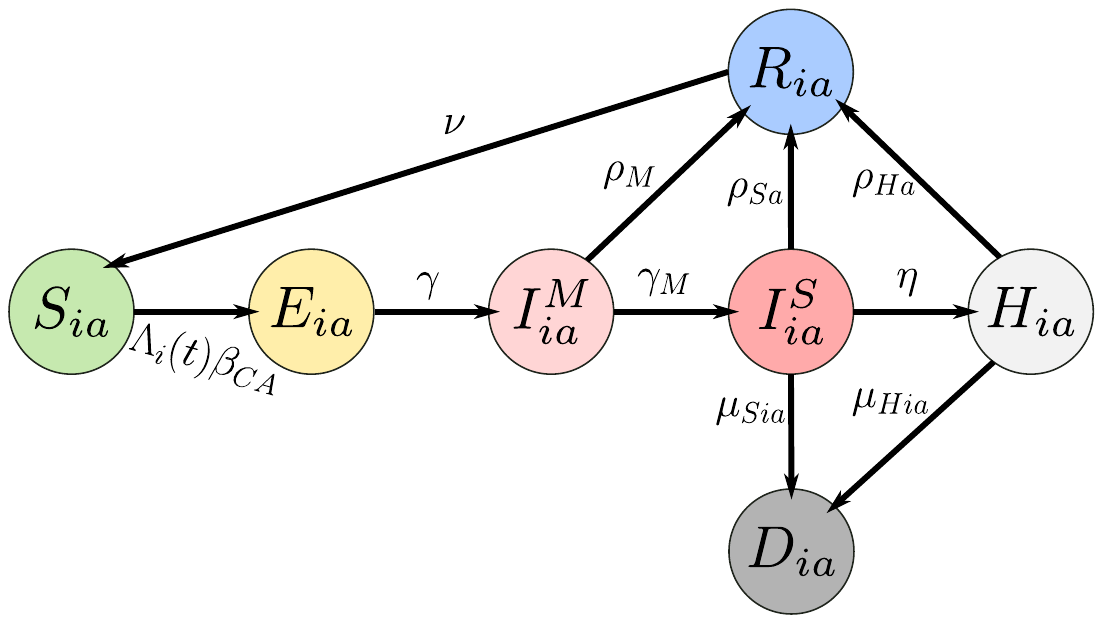}
  \caption{Schematic of infection stages in SCoVMod. Individuals pass
    through stages post infection as described by arrows. Not all
    stages are obligatory for all infected individuals (e.g. some
    individuals recover without going to hospital).}
  \label{fig:model-structure}
\end{figure}

Within each OA ($i$) the infection process is governed by a compartmental model (Figure~\ref{fig:model-structure}) for which the frequency dependent force of infection $\Lambda _i(t)$ defined in Figure~\ref{fig:force}. In the compartmental model are infection classes $S$ (susceptible), $E$ (exposed), $I^M$ (mildly infected), $I^S$ (severely infected), $H$ (hospitalised).

The state transitions in the model are described by the following equations:

\begin{align*}
  \dfrac{dS_{ia}}{\mathit{dt}}& =-\Lambda_i(t) \beta_{CA} S_{ia} + \nu R_{ia}\\
  \dfrac{dE_{ia}}{\mathit{dt}}& =\Lambda_i(t) \beta_{CA} S_{ia} - \gamma E_{ia}\\
  \dfrac{dI_{ia}^M}{\mathit{dt}}& =\gamma E_{ia} - (\gamma_M + \rho_M) I_{ia}^{M}\\
  \dfrac{dI_{ia}^S}{\mathit{dt}} & =\gamma_{M} I_{ia}^{M} - (\rho_{Sa} + \mu_{Sia} + \eta) I_{ia}^{S}\\
  \dfrac{dH_{ia}}{\mathit{dt}} & =\eta I_{ia}^{S} - ( \rho_{Ha} + \mu_{Hia}) I_{ia}^{S}\\
  \dfrac{dR_{ia}}{\mathit{dt}} & =\rho_{M} I_{ia}^{M} + \rho_{Sa} I_{ia}^{S} + \rho_{Ha} H_{ia} - \nu R_{ia}\\
  D_{ia} & = N_{ia} - ( E_{ia} + I^{M}_{ia} + I^{S}_{ia} + H_{ia} + R_{ia} )
\end{align*}

The force of infection, $\Lambda$, is detailed in Figure~\ref{fig:force} and other state transition rates are given by: $\gamma$ for $E \rightarrow I^M$, $\gamma_M$ for $I^M \rightarrow I^S$, $\eta$ for $I^S \rightarrow H$, $\rho_M$ for $I^M \rightarrow R$, $\rho_{Sa}$ for $I^S\rightarrow R$  for age class $a$, $\rho_{Ha}$ for $H \rightarrow R$ for age class $a$, $\mu_{Sia}$ for $I^S \rightarrow D$ for age class $a$ and location $i$, $\mu_{Hia}$ for $H \rightarrow D$ for age class $a$ and location $i$, and $\nu$ for $R \rightarrow S$.

Transmission rates are adjusted by location, according to SIMD Health Index:
\begin{equation*}
  \beta_{CA} = 1 + \left(\beta_{mod} (k_{CA}-k_{av}) \right)
\end{equation*}
where $\beta_{CA}$ is the transmission modifier rate for a given CA, $k_{CA}$ is the CA mean health index value (from the SIMD), and $\beta_{mod}$ is a fitted parameter for the strength of the overall effect.

For this study the model is extended with a second strain (variant) of the virus. The compartment structure remains the same, but individuals can be infected with one of two strains, each with a different set of rates. We assume complete cross immunity between strains.

The values for all parameters are either established from the literature (Table~\ref{tab:params}) or fit (see below).

\subsubsection*{National transmission (between-OA movements)}
Between OAs individuals move daily across a network of locations defined by Scottish Census data adjusted by Google mobility data. 

From the current population estimates we draw the number of individuals whose primary residence is mapped onto an OA, with their age group. The total population of Scotland from this estimate is 5,438,054 (Young: 919,580; Adult: 3,492,421; Elderly: 1,026,053). Of the adults 1,960,712 commute to work, reduced to 647,034 under lockdown (see details below).

An individual's workplace is assigned by distributing a proportion of the population of each location to each work location, weighted by the proportion of individuals from each home location in the census flows data who work in another location.  For the remaining proportion we synthetically generated daytime locations by randomly selecting OAs either from the OAs within the same intermediate zone (a geographical area containing approximately 4200 people), with probability 0.9, or from the OAs within a neighbouring intermediate zone, with probability 0.1.

For each origin $o$ and destination $d$ we assign a weight $w_{od}$ from the census flow data:
\[ w_{od} = \frac{n_{od}}{t_o} \]
where $n_{od}$ is the total number of people who move from $o$ to $d$ to work, and $t_o$ is the total number who move from origin $o$ to any location for work. We take the individuals of each home location if they are eligible to work (total $n_o$); in this case we assume all individuals of adult age 16–64. Each destination is assigned to $n_o \times w_{od}$ of these individuals. The individuals who remain have no assigned workplace---we assume either they do not work, or they work within their home location.

For each day of the simulation we consider two time steps: a day step where individuals can move to their place of work, and a night step where those individuals move back to their home location. In each day step, we take each destination location $d$. Let $\lambda_d$ be the number of eligible workers who may move to the destination location. The number of moves $s$ is then scaled according to the per cent change in mobility $m$ (see below) for the given day: $s_m = \lfloor s(1 + \frac{m}{100}) \rfloor$.

In order to improve the computational efficiency of the overall simulation, movements of commuters between OAs were batched into groups of 5, with movements between OAs of fewer than five individuals per day retained at a proportionate rate by drawing from a binomial distribution: $s_{mt}\sim B(s_{m},\frac{1}{5})$. If the sampled number of workers $s_{mt}$ is less than or equal to the number of workers who may normally move to destination $d$, then those who move are sampled randomly from those who may normally move. However, if $s_{mt}$ is greater than the number of workers who may normally move to $d$, then the additional workers are drawn randomly from workers who have no assigned destination location. While this reduces the overall network link density, the effect on transmission dynamics in this setting is negligible. We note that this means that interpretation of the combined $\beta_D$ and $\beta_N$ must be made with caution and not compared directly to other models.

For each night of the simulation, the workers who moved in the day step are moved back to their origin location.

\subsubsection*{Vaccination}
Vaccination is represented in the model by flagging an individual with its vaccination status. The transmission rates affecting a vaccinated individual are adjusted accordingly---i.e.\ the probability that a vaccinated individual receives an infection is reduced according to vaccine effectiveness. Vaccination numbers per DZ in this study were initiated as recorded in eDRIS data, and all doses and booster doses were applied to individuals in the model, within each DZ and each age class, in the same proportions as were administered in the real population. Vaccine effectiveness in each dose phase is a fixed parameter per phase and per strain.

For the period after 11th Dec 2021, we assumed a modelled distribution of the booster doses that were being delivered at that time. We distributed booster doses to all individuals in the model, who have had their second dose, after three months (as per the actual distribution schedule at the time). We ran one scenario with 100\% uptake w.r.t.\ 2nd dose, and one with 55\% uptake. At a national level 2097633 boosters had been administered by 11 Dec 2021, 3814877 second doses had been administered by 18 Sep 2021 (12 weeks prior, thus eligible for a booster on 11 Dec 2021), thus 55\% of all eligible had had their booster by 11 Dec.

More recently we also ran a final scenario based on the recorded vaccine distribution beyond the 11th December for comparison.

Based on the available information at the time~\cite{andrews2021} and consistent with other approaches~\cite{barnard2021, keeling2021}, we assume no reduction in outcome severity associated with vaccinated individuals, but rely on the transitive effect of reducing associated transmission.

\subsubsection*{Modelling COVID-19 incidence}
As PHS eDRIS data gives us only the reported cases of infection, we need to determine the likely number of infected individuals with which to seed the model and to use as the summary statistic for validation and parameter estimation. For this we using methods described in Colman et al.~\cite{colman2023} we estimated the proportion of infections ascertained to be $a=0.25$ and thus the likely number of infected individuals on a given day, $\tau$, to be  $I_{\tau}=C_{\tau}/a$ where $C_{\tau}$ is the number of reported cases on day $\tau$. As case ascertainment is expected to vary by region, we use testing data for the $32$ subregions of Scotland, known as council areas, to improve our estimates of incidence. Supposing the number of tests in the region is $s_{\tau}$, and $c_{\tau}$ of them are positive, while at the national scale the number of tests is $S_{\tau}$ and $I_{\tau}$ are positive, and finally the population of the region is $n$, the formula
\begin{equation}
    \label{mu}
    \mu_{t}=\frac{\displaystyle \sum_{i=1}^{t}(s_{i}+c_{i})z^{t-i}}{\displaystyle \sum_{i=1}^{t}(s_{i}+nS_{i}/I_{i})z^{t-i}}
\end{equation}
to give the expected number of individuals in the region who would potentially test positive (if tested) on day $t$. Here $z$ is a hyperparameter that we are able to adjust between smooth ($z>0$) or more responsive ($z\approx0$) outputs. We choose $z=0.1$. The derivation of this formula is given in the supplementary information.

We convert this daily estimate of the number of people testing positive into a weekly number of new infections by summing $\mu_{t}$ for each day across the $7$-day period, then multiplying by a constant to account for the duration of test-sensitivity and to correct for infected individuals who are never test-sensitive. The value of the constant, $0.265$, was chosen in such a way to create agreement between the obtained incidence at the national scale (by summing across all regions) and other estimates derived using the method described in Colman et al~\cite{colman2023}.

\subsubsection*{Modelling transmission rate changes over time}
To model the changes in activity over time---e.g.\ those that are the effect of lockdowns, but also other more voluntary changes in behaviour---we consider two factors. First, we thin movements in the simulation (mobility reduction) in proportion to observed changes in mobility according to Google mobility reports~\cite{googleinc.2022}. This is applied as a proportional change in the number of individuals making between-OA movements on a given day (as above).

Second, physical distancing is incorporated via a reduction in contacts applied to both daytime and nighttime transmission rates (transmission reduction). Beyond the initial fit period (see below), we fit a change in transmission rate over time only, assuming that posterior distributions for all other parameters estimated based on the initial fit remain relevant.

\subsubsection*{Parameter estimation}
Simulated epidemics are compared to the spatio-temporal pattern of COVID-19 spread in Scotland. Non-observable parameters were estimated using the incidence of COVID-19, according to the model described above, during the initial Delta-dominant period.

Estimation was performed using a Sequential Monte Carlo implementation of Approximate Bayesian Computation (ABC-SMC)~\cite{Hartig2011,Toni2009}. We calibrated the model output to the weekly incidence (number of estimated infections per week) due to COVID-19 aggregated at the level of CAs, using this spatial variation in incidence across Scotland to provide the necessary signature to properly calibrate the role of human mobility.

Simulated and observed summary statistics are compared via a score equal to a sum of squared errors, recorded weekly:
\[
  \mathit{score} = \sum_{w} \sum_{l} \left(I_{sim}(w,l)-I_{obs}(w,l)\right)^2
\]
where $I_{sim}(w,l)$ is the weekly ($w$) incidence per CA ($l$) simulated and $I_{obs}(w,l)$ its observed value.

The total number of infected individuals at the start of the simulation (the seeds) are fitted as part of the inference. The seeds are randomly assigned a disease state from $E$, $I^M$, and $I^S$. Seed locations are distributed according to the proportion of infections registered per Intermediate Zone on the date of the start of the simulation. Intermediate zones are \emph{neighbourhood level} aggregates of approximately five DZs.

Uniform prior distributions constrain all parameter values to plausible ranges based on the available literature relevant to the early, pre-lockdown period. Infection dynamics are simulated via a $\tau$-leap algorithm using half-day timesteps~\cite{Gillespie2001}. All parameters are listed in Table~\ref{tab:params}.

The inference framework is run on a distributed application framework (Akka)~\cite{lightbendinc.2022} running on a cloud computing infrastructure (Amazon AWS2)~\cite{amazonwebserivesinc.2022}. The model code has been written using industry grade software engineering practices including agile development for project task planning, test driven development, pair programming and code reviews to produce unit tested, robust, and reusable software components. The majority of the code has been reviewed by at least one other software developer and the source code is available\footnote{SCoVMod Omicron source code: \url{https://github.com/Kao-Group/SCoVMod-Omicron}}.

\subsubsection*{Fitting transmission rate changes over time}
After fitting the initial parameter set, a further temporal refinement is employed to improve the fit where the model cannot account for external factors such as changes in transmission rate owing to NPIs or other changes in human behaviour not captured by the mobility data. Here we perform a piece-wise least-squares fit over just the transmission rate $\beta$, per Council Area. This is piece-wise in the inflection points in the case data, which largely correspond to the times at which NPIs were enacted or relaxed.

\subsection{Scoping scenarios}
We modelled a range of scenarios, considering two variants (Delta and Omicron) with differing transmission rate advantage and levels of vaccine protection. At the time of the analysis, there were limited data on the potential of the Omicron VOC for either greater transmission rates than the Delta VOC, or increased ability to escape either natural or vaccine-induced immunity. We therefore generated simulations over possible transmission rates and vaccine escape levels for the Omicron variant, based on this existing evidence, in order to choose combinations that generate plausible trajectories of SGTF cases in Scotland. We used values for transmission rate advantage and vaccine efficacy from the UKHSA Technical Briefing on Variants of Concern~\cite{ukhealthsecurityagency2021}.

We modelled three levels of vaccine escape: a baseline with the same vaccine efficacy as Delta (Escape Level 1), a lower vaccine escape potential (Escape Level 2) as described by the central estimates from the UKHSA Technical Briefing (90\% after two doses, falling to 35\% after 15 weeks, and 75\% after booster), and a higher vaccine escape potential (Escape Level 3) as described by the lower efficacy bounds (65\% after two doses, falling to 10\% after 15 weeks, and 60\% after booster).

We modelled two levels of transmission rate advantage for Omicron: an increased level based on the UKHSA Technical Briefing estimate of 3.2$\times$ household transmission, under the assumption that generalised transmission rate advantages, i.e. including transmission outside the household has a similar advantage, and an intermediate level based on 80\% of the higher level, giving a 2.25$\times$ transmission rate---corresponding with the lower bound of the UKHSA estimate. We then have two sets of scenarios. The first assumes the overall transmission rate for Delta remains at the current level. The second assumes an NPI-based intervention, reducing transmission rates to 80\% its previous value on 17th December 2021, to reflect a combination of voluntary behavioural change in response to new guidance, and new restrictions. The latter scenario aims to generate a reproduction number around 0.8 for the Delta VOC, as observed during previous similar NPI restricted periods in Scotland~\cite{scottishgovernment2022b}. 

We assumed that all parameters for the Omicron VOC were the same as for Delta, with the exception of the transmission rate, which was assumed to be a fixed multiplier of the transmission rates for Delta. The transmission rates for Delta are from parameters jointly drawn from the posterior parameter distributions fitted for Delta from the earlier period (i.e. from August 2021 to November 2021). Vaccine escape levels were implemented as a multiplier on the within-model vaccine efficacy for Delta as in Table \ref{tab:VaccEscapeScen}. The combination of vaccine escape and transmission parameters generates 12 scenarios (a--l), labelled as in Table \ref{tab:scenarios}.

\begin{table}
  \centering
  \begin{tabular}{|l l|l|l|l|}
    \hline
    \multicolumn{5}{|r|}{Efficacy multiplier}\\
    & & 2 dose & +15 weeks & Booster \\
    \hline
    \multirow{3}{*}{Vaccine Escape Level} & 1 & 1 & 1 & 1 \\
    & 2 & 1 & 0.45 & 0.8 \\
    & 3 & 0.72 & 0.15 & 0.63 \\
    \hline
  \end{tabular}
  \caption{\label{tab:VaccEscapeScen} Levels of vaccine escape (1,2,3) used in our scenarios, with level 1 being complete protection (an efficacy multiplier of 1 on all doses). Efficacy multiplier is applied to the base efficacy level for the vaccine dose.}
\end{table}

The transmission model does not take account of vaccine protection from severe disease. It either completely protects individuals or fails completely and so if infected, results in an infection as likely to be severe as would be the case for infection in a previously wholly susceptible individual. 

\begin{table}
  \centering
  \begin{tabular}{|l l|l l l|}
    \hline
    \multicolumn{5}{|r|}{Vaccine Escape Level}\\
    & & 1 & 2 & 3 \\
    \hline
    \multirow{4}{*}{Transmission Level}
    & 1 & a & b & c \\
    & 2 & d & e & f \\
    & 1+NPI & g & h & i \\
    & 2+NPI & j & k & l \\
    \hline
  \end{tabular}
  \caption{\label{tab:scenarios} Scenario labels as used in figures in the exploration of scenario plausibility.}
\end{table}

\subsection{Fitting of plausible scenarios}
From the scoping scenarios, we carried forward only those that were plausible, being closest to the observed SGTF incidence trajectory. We then adjusted the transmission rates of each to fit the growth rate observed in incidence.
\[
    \text{Growth rate} = \frac{\ln I(t_f) - \ln I(t_0)}{t_f - t_0}
\]
where $I(t)$ is the number of infected at time $t$ from an initial point ($t=t_0$) fitted to match the SGTF incidence trajectory. We use the time period for which we have confirmed observed Omicron cases. We then take the growth rates for the modelled scenario ($g_m$) and for the observed incidence ($g_o$) and increase the scenario transmission rate by $\frac{g_o}{g_m}$. The reported transmission rate advantage is expressed as this value  $\frac{g_o}{g_m}$.

We then took a range of further scenarios:

\vspace{1em}
\begin{tabular}{|l | l l l l|}
  \hline
  &\multicolumn{4}{|r|}{NPI Level}\\
  & 1 & 2 & 3 & 4 \\
  \hline
  Lower vaccine escape & m & o & q & s \\
  Higher vaccine escape & n & p & r & t \\
  \hline
\end{tabular}
\vspace{1em}

\noindent where NPI Level 1 is as considered previously, an initial drop to 80\% of the original transmission rate. Levels~2,~3,~and~4 have a further post-Christmas restriction applied with a further drop to 64\%, 48\%, and 40\% of the original transmission rates respectively, imposing these restrictions on 27th December 2021. As the mean of the fitted trajectories of the fitted Delta period were observed to stabilize after 100 runs, further simulations were restricted to this number.

We take the value $\frac{g_o}{g_m}$ as the estimated transmission rate multiplier for the Omicron VOC, for both the lower and higher vaccine escape scenarios.

\subsection{Hospital occupancy estimates}
We estimate trajectories of hospital admissions, and overall hospital occupancy, from the model output infections. Here, we assume that that the age distribution of infections, rates of hospital admissions and length of hospital occupancy are identical to that observed over the period 1 May 2021 -- 1 December 2021, during which the Delta variant was dominant. 

Methods are described in detail by Wood and Kao~\cite{wood2022}. We estimate a trajectory of hospital admissions and hospital occupancy by time \emph{convolutions} -- transformations that delay and ``draw out'' the trajectories based on known variation in the time between cases and hospitalisations, and in the time spent by individuals in hospital with COVID-19. The admissions trajectory is first estimated by a convolution of cases with the distribution of times between cases and hospitalisations, drawing from empirical distributions obtained in this prior period. We then estimate an \emph{occupancy} trajectory by a convolution of the admissions trajectory with the distribution of time spent by individuals in hospital with COVID-19. The number of cases is scaled relative to infections by a case ascertainment of 25\% (an estimate consistent with historical case ascertainment in the UK when testing was widely available~\cite{colman2023}). The overall hospitalisation rate (the proportion of cases that result in an admission) is estimated at 2.5\%, age-stratified based on hospital admissions in the prior period (Table~\ref{tab:hosprates}).

\section{Results}
\subsection{Model fit / parameters}
\begin{figure}
  \centering
  \includegraphics[width=0.49\textwidth]{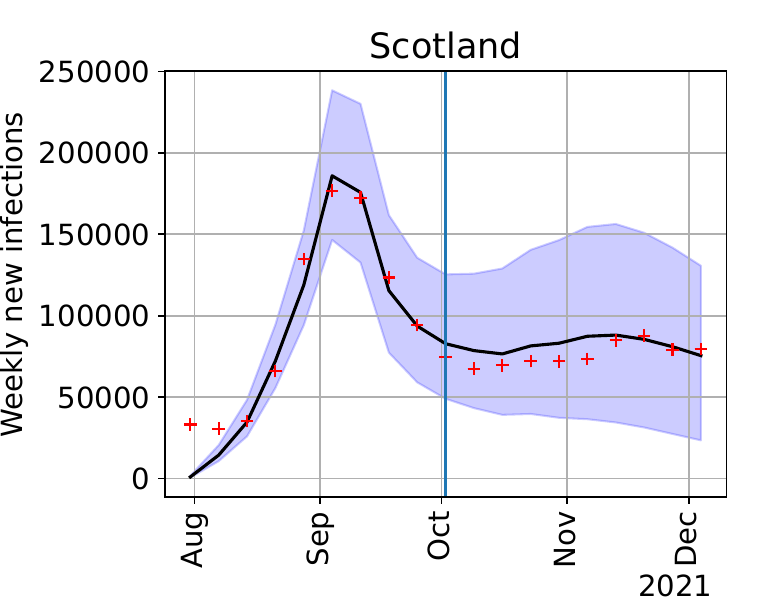}
  \caption{Model fit (black line, blue confidence intervals) to observed incidence (red crosses) for the Delta variant, after the inital ABC-based parameter estimation and temporal transmission rate fit, national scale. The vertical line shows the final inflection point in the piece-wise temporal fit.}
  \label{fig:fit-diagnostic-nat}
\end{figure}

\begin{figure}[h]
  \centering
  \includegraphics[width=\textwidth]{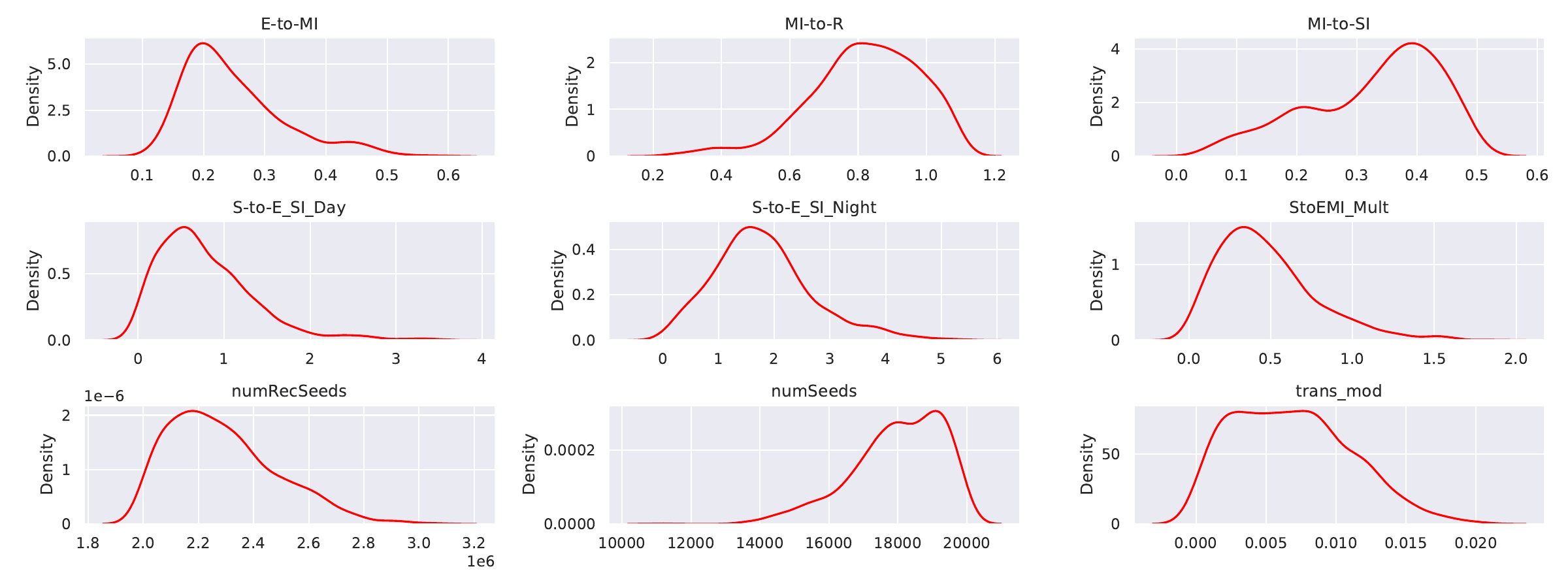}
  \caption{Posterior distribution of parameters, as a result of the initial ABC-based parameter estimation.}
  \label{fig:posterior}
\end{figure}

\begin{figure}[h]
  \centering
  \includegraphics[width=\textwidth]{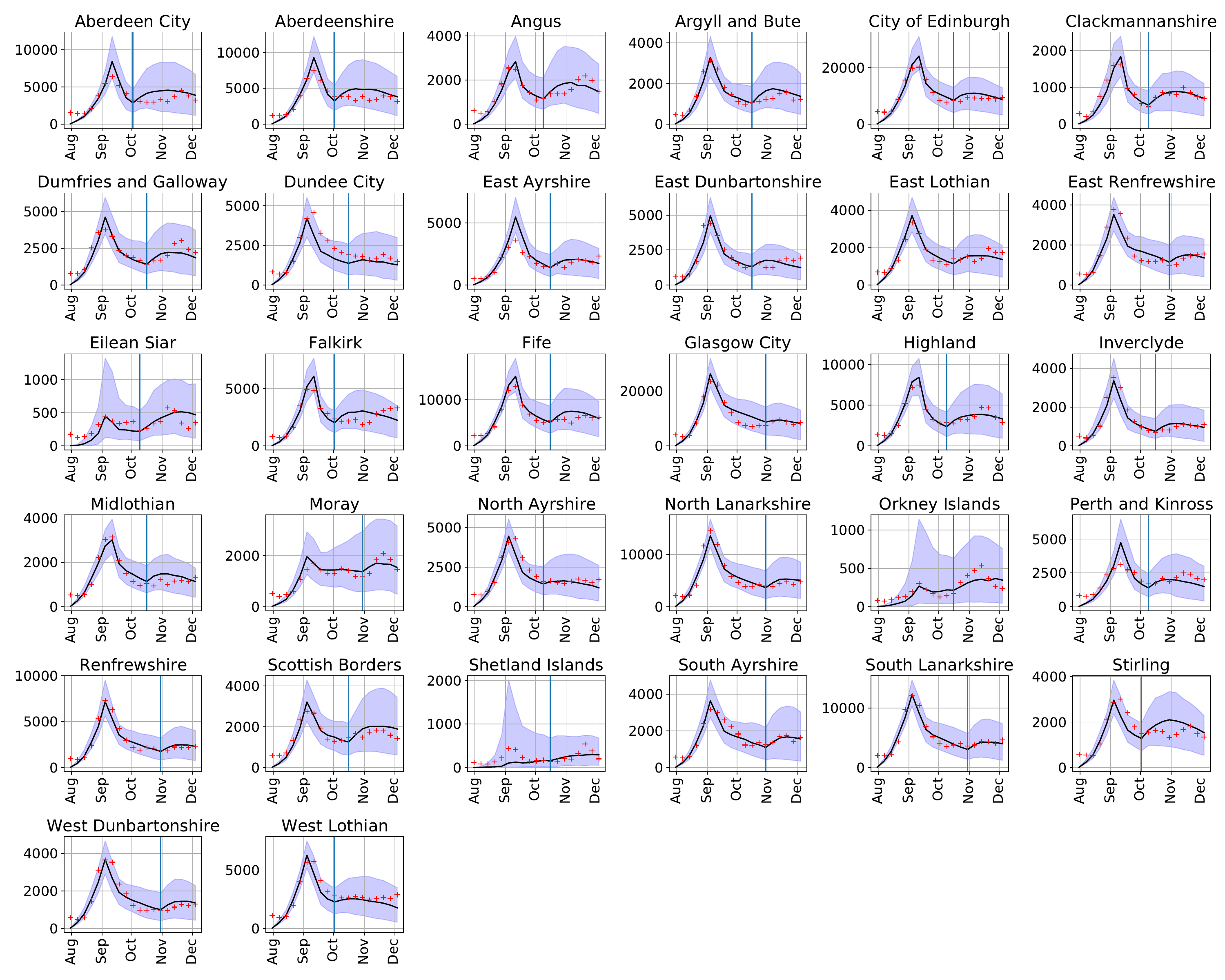}
  \caption{Council Area breakdown of the model fit to observed data, after the inital ABC-based parameter estimation and temporal transmission rate fit.}
  \label{fig:fit-diagnostic-CA}
\end{figure}

The fit of the initial ABC-based parameter estimation to the observed data, with the resultant posterior distribution of estimated parameters is shown in Figure~\ref{fig:posterior}. Figure~\ref{fig:fit-diagnostic-nat} shows the trajectory of incidence in the model compared to the observed data (accounting for case ascertainment), with detail at the Council Area level shown in Figure~\ref{fig:fit-diagnostic-CA}. Notably, the  final imposed transmission rate change in the piece-wise temporal transmission rate fit is at the start of October 2021, with the final incidence inflection point in mid-November being a natural result of the modelled dynamics and fitting well with the observed incidence.

\subsection{Scoping scenario results}
\begin{figure}
  \centering
  \includegraphics[width=0.49\textwidth]{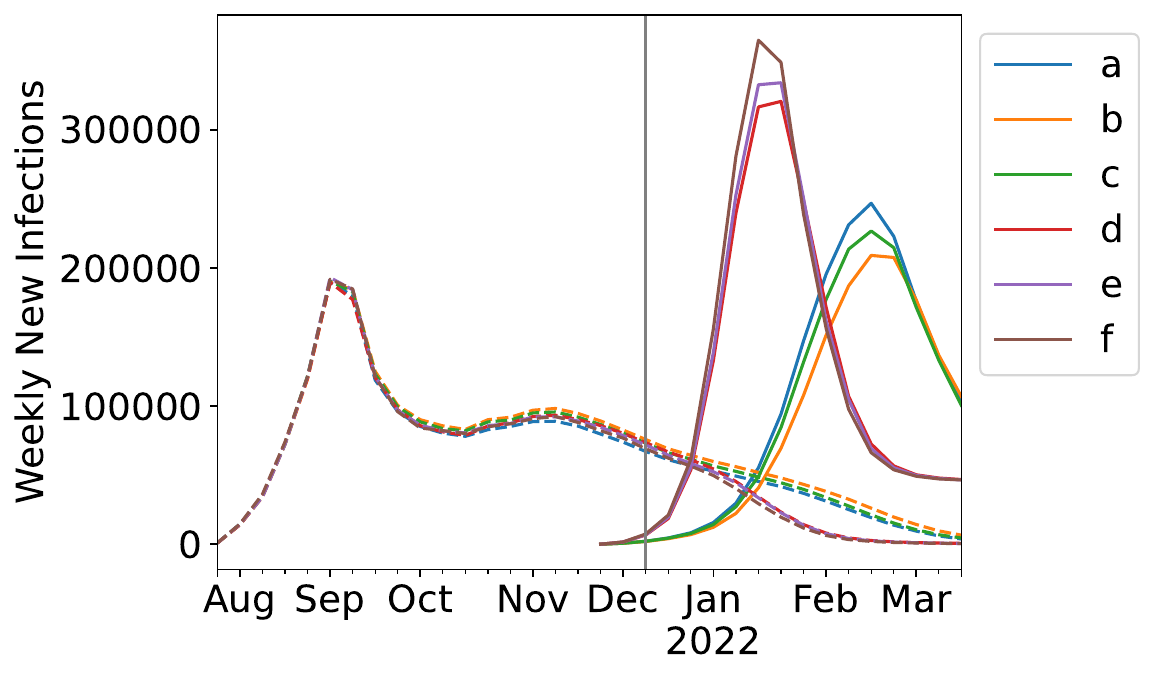}
  \includegraphics[width=0.49\textwidth]{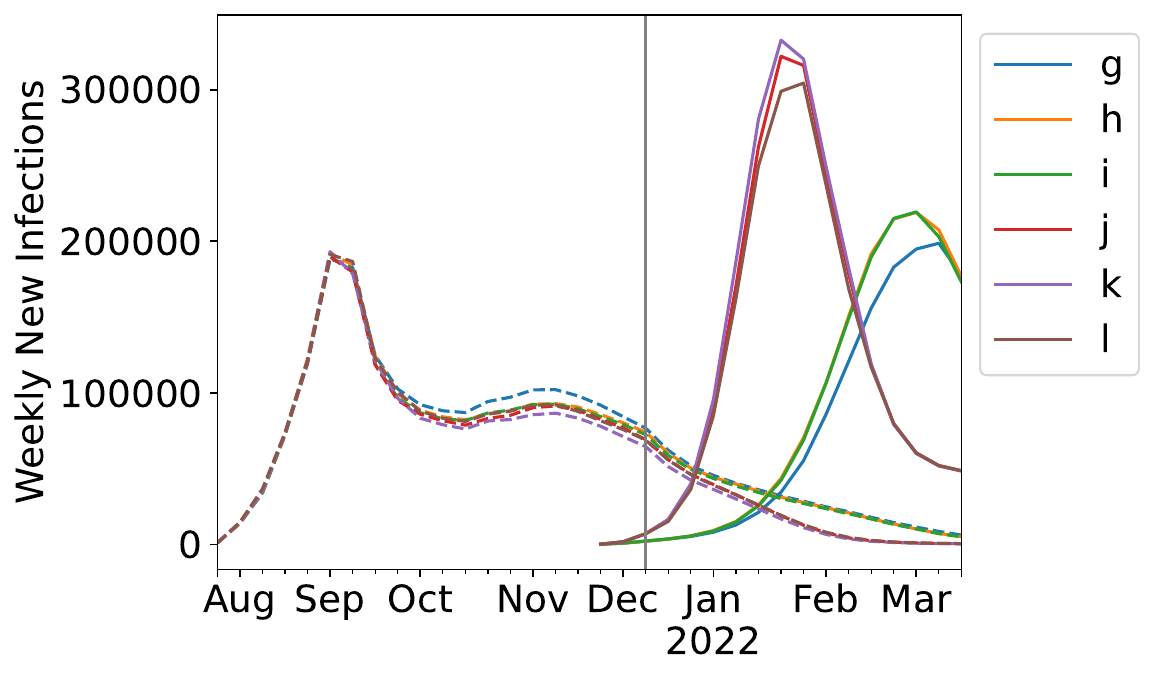}
  \caption{Comparison of scenarios with no additional NPIs (left) and additional NPIs (right). Central estimate from 200 model runs, from the vaccination distribution scheme with 55\% uptake. With lower transmission rate: a--c, g--i; higher transmission rate: d--f, j--l; and increasing vaccine escape within each group. The trajectory of the Delta variant is shown as a dashed line. Simulation trajectories with confidence intervals are shown in Figures~\ref{fig:55:scenario-grid-1}~and~\ref{fig:55:scenario-grid-2}.}
  \label{fig:scenario-comp}
\end{figure}

The dynamics of infection with respect to the Delta and Omicron variants are directly compared in Figure~\ref{fig:scenario-comp} showing trajectories both with and without NPI's in place.

\begin{figure}[h]
  \centering
  \includegraphics[width=\textwidth]{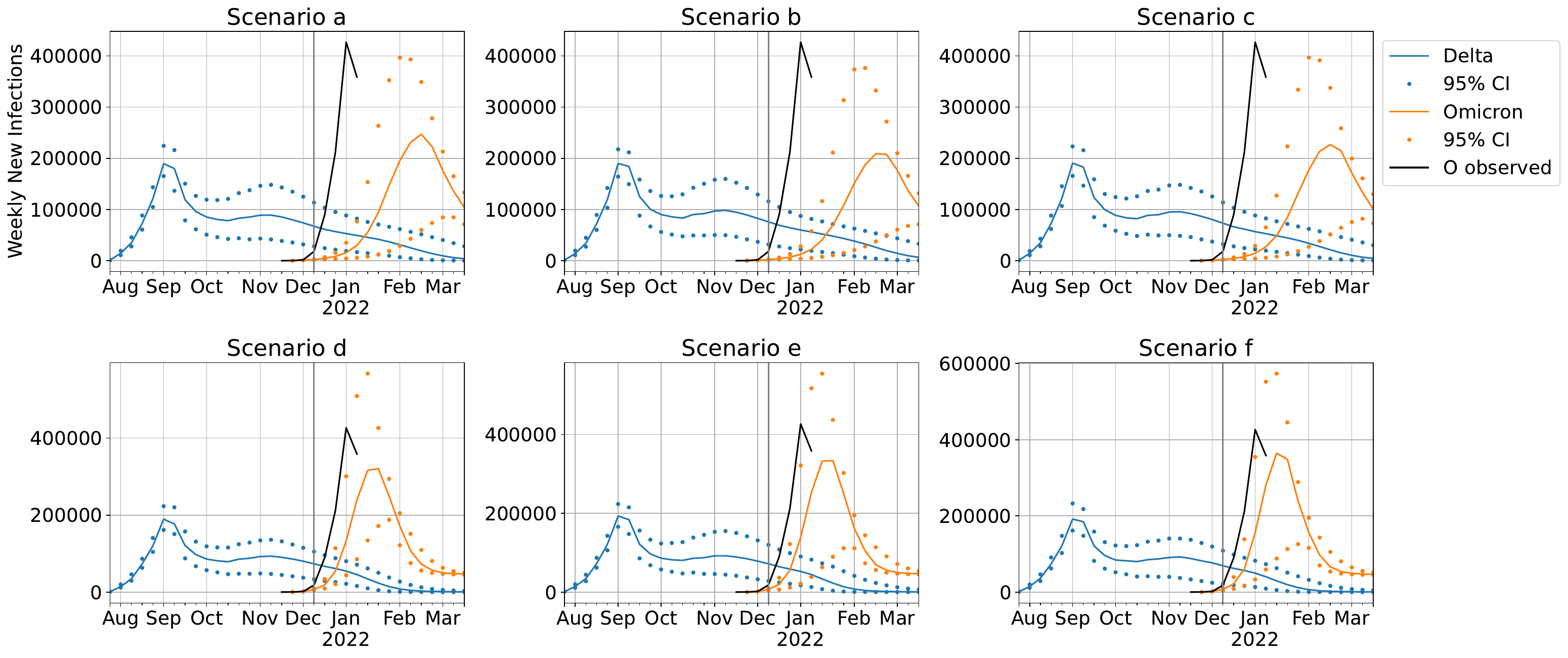}
  \caption{Scoping scenarios with increasing vaccine escape from left to right, increasing Omicron transmission rate from top to bottom. 55\% uptake vaccine distribution scheme.}
  \label{fig:55:scenario-grid-1}
\end{figure}

\begin{figure}[h]
  \centering
  \includegraphics[width=\textwidth]{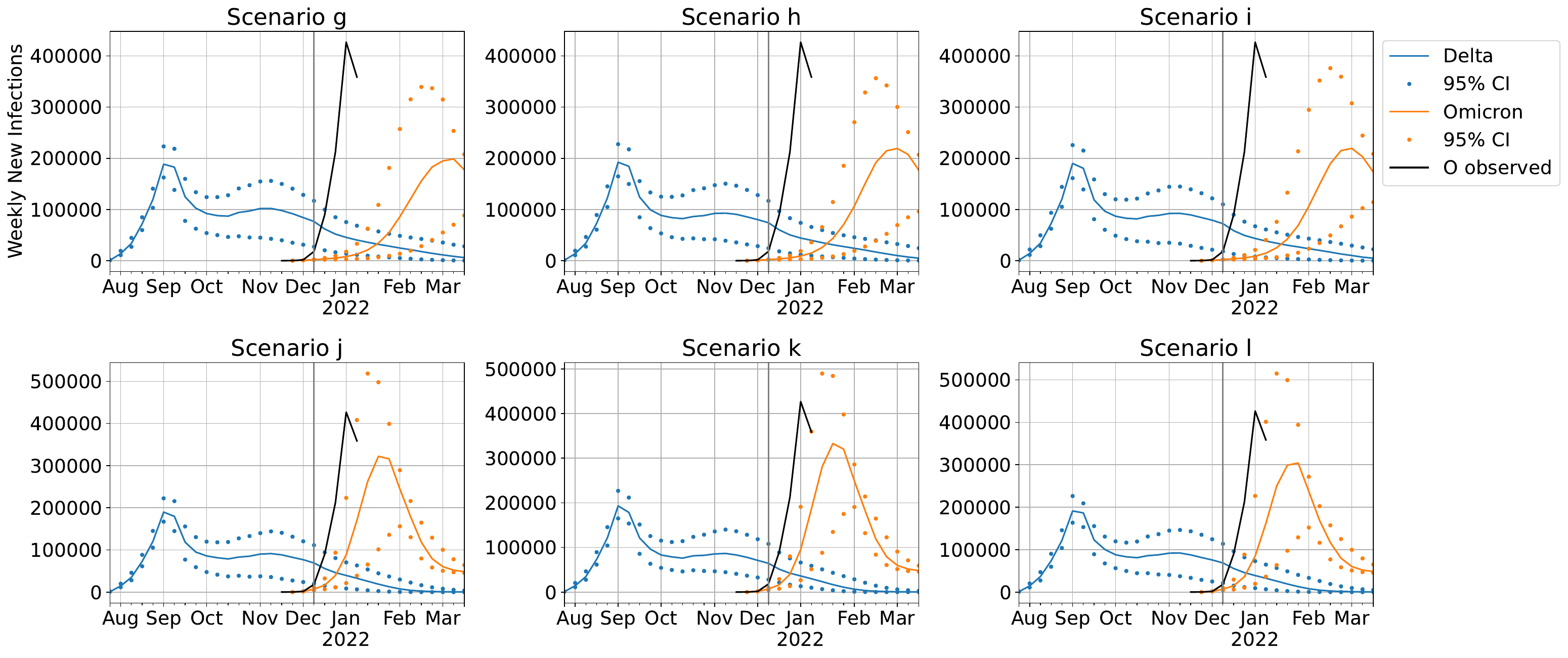}
  \caption{Scoping scenarios with additional NPI reducing transmission rates to 80\%: increasing vaccine escape from left to right, increasing Omicron transmission rate from top to bottom.  55\% uptake vaccine distribution scheme.}
  \label{fig:55:scenario-grid-2}
\end{figure}

Figure~\ref{fig:55:scenario-grid-1} and Figure~\ref{fig:55:scenario-grid-2} provide additional detail, including confidence intervals, showing respectively trajectories without and with further NPI's in place from 17th December, and showing the range of simulations and a comparison to the observed incidence. 

\subsection{Selection of plausible scenarios}
\begin{figure}[h]
  \centering
  \includegraphics[width=\textwidth]{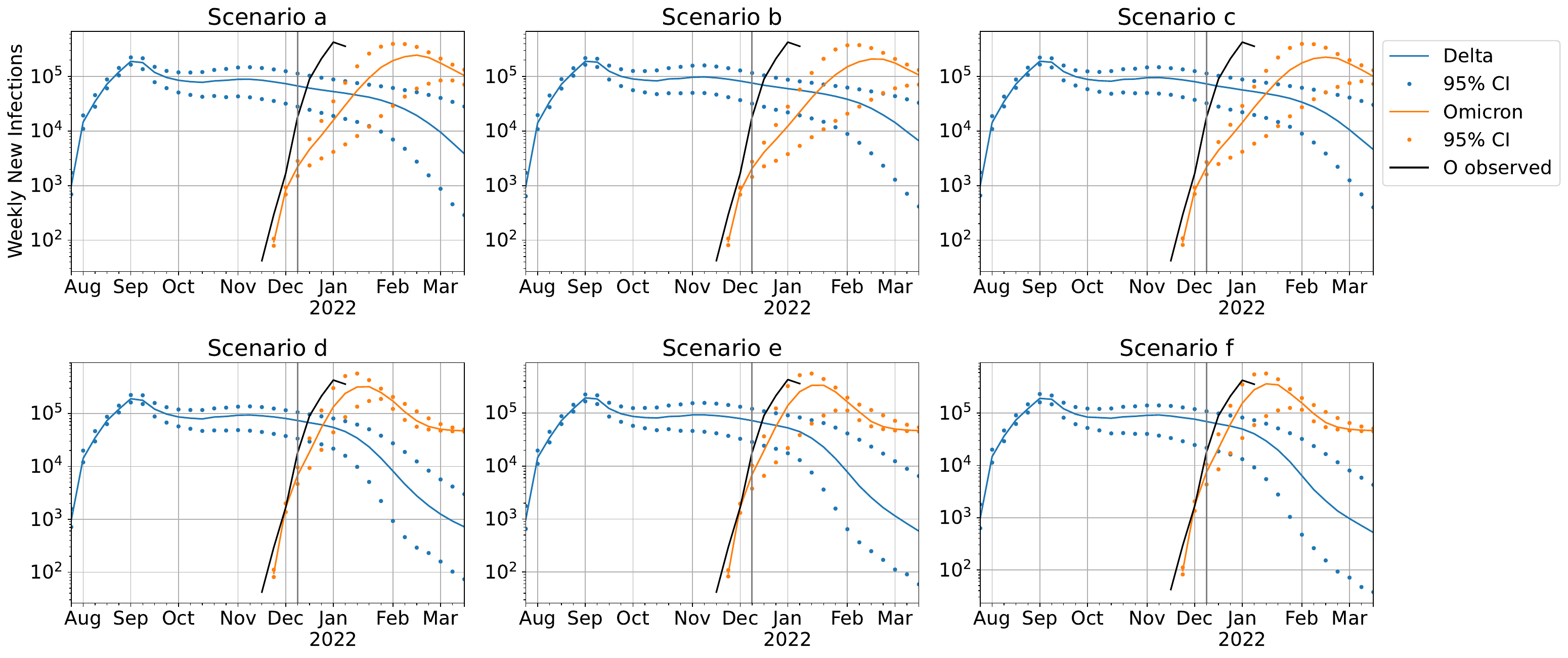}
  \caption{Scoping scenarios with increasing vaccine escape from left to right, increasing Omicron transmission rate from top to bottom. 55\% uptake vaccine distribution scheme. (Log scale.)}
  \label{fig:55:scenario-grid-1-log}
\end{figure}

\begin{figure}[h]
  \centering
  \includegraphics[width=\textwidth]{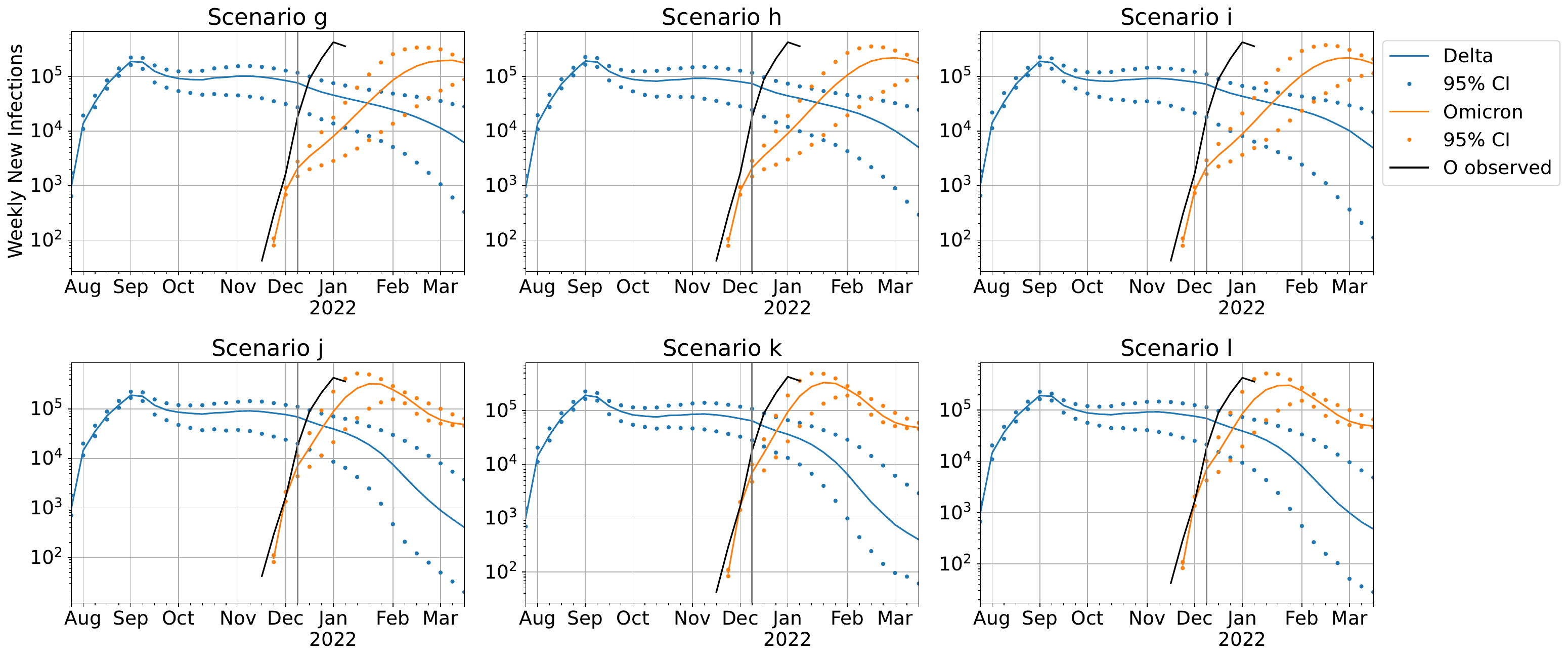}
  \caption{Scoping scenarios with additional NPI reducing transmission rates to 80\%: increasing vaccine escape from left to right, increasing Omicron transmission rate from top to bottom.  55\% uptake vaccine distribution scheme. (Log scale.)}
  \label{fig:55:scenario-grid-2-log}
\end{figure}

We compared scenarios to the observed incidence of SGTF-only cases adjusted by our modelled case to infection ratio. Only a few scenarios, all with the higher transmission rate were close to estimating the true trajectory (growth rate) of Omicron infections (Figures~\ref{fig:55:scenario-grid-1-log} and~\ref{fig:55:scenario-grid-2-log}). Hence, we restricted further scenario development to the most plausible of these.

We carry forward only scenarios k and l as being closest to the observed SGTF case trajectory, also under the assumption that the measures already in place would have had some impact. Scenario k has the lower and l the higher estimate for vaccine escape. Both have the higher transmission rate estimate. We then adjust the transmission rates of each to fit the growth rate observed in cases.

\subsection{Fit scenario results}
\begin{figure}
  \centering
  \includegraphics[width=\textwidth]{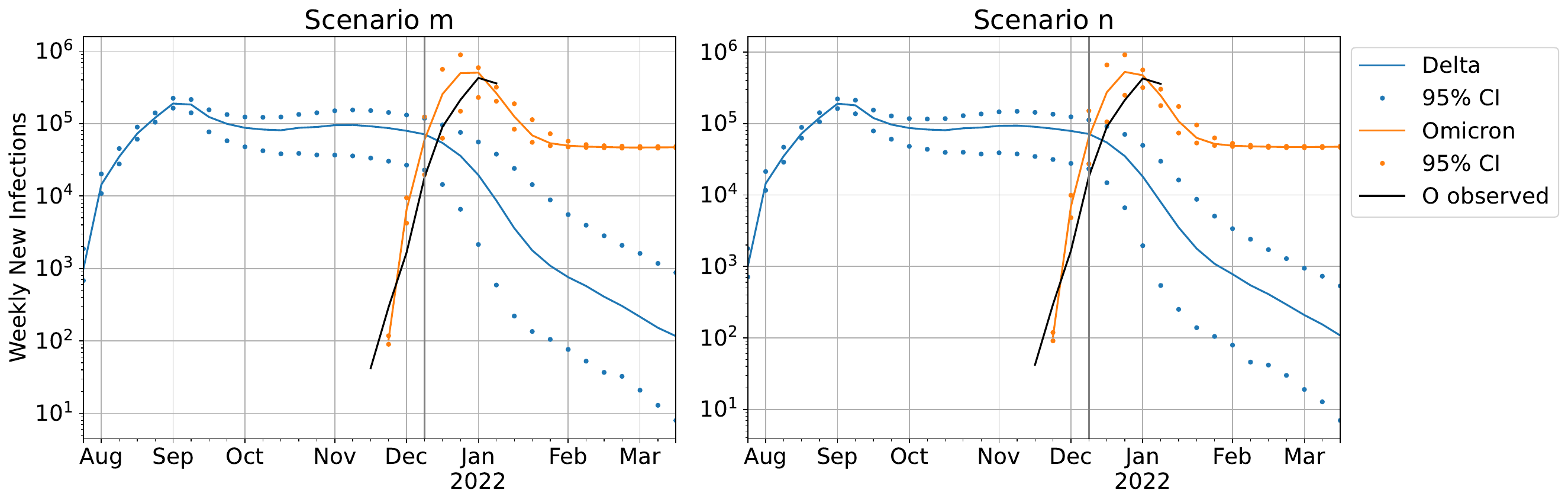}
  \caption{Adjusted transmission rate scenarios (log scale) for the 55\% uptake vaccine distribution scheme: lower vaccine escape (left), higher vaccine escape (right). Dots represents bounds for 95\% of 200 simulations. Observed values prior to the vertical line (11th December) were the data available at the time of fitting, used to fit the growth rate; observed values subsequent to this date were added later for comparison.}
  \label{fig:55:scenario-grid-mn}
\end{figure}

By fitting the growth rate to that of the observed Omicron cases, we achieved an initial modelled growth rate consistent with the limited knowledge of the early Omicron outbreak. Figure~\ref{fig:55:scenario-grid-mn} shows the fit to observed case data (including knowledge of observed Omicron cases after 11th December that were not known at the time of original modelling).

\begin{figure}[h]
  \centering
  55\% uptake distribution:\\
  \includegraphics[width=0.49\textwidth]{scenario_grid_mn_log}
  \includegraphics[width=0.49\textwidth]{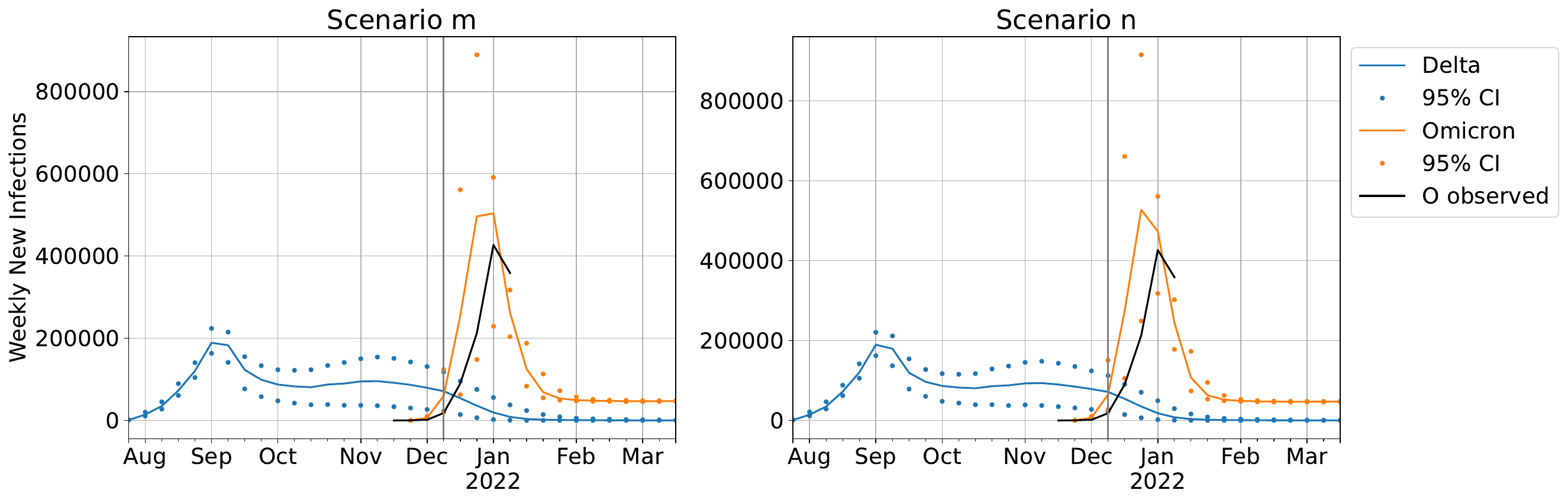}\\
  100\% uptake distribution:\\
  \includegraphics[width=0.49\textwidth]{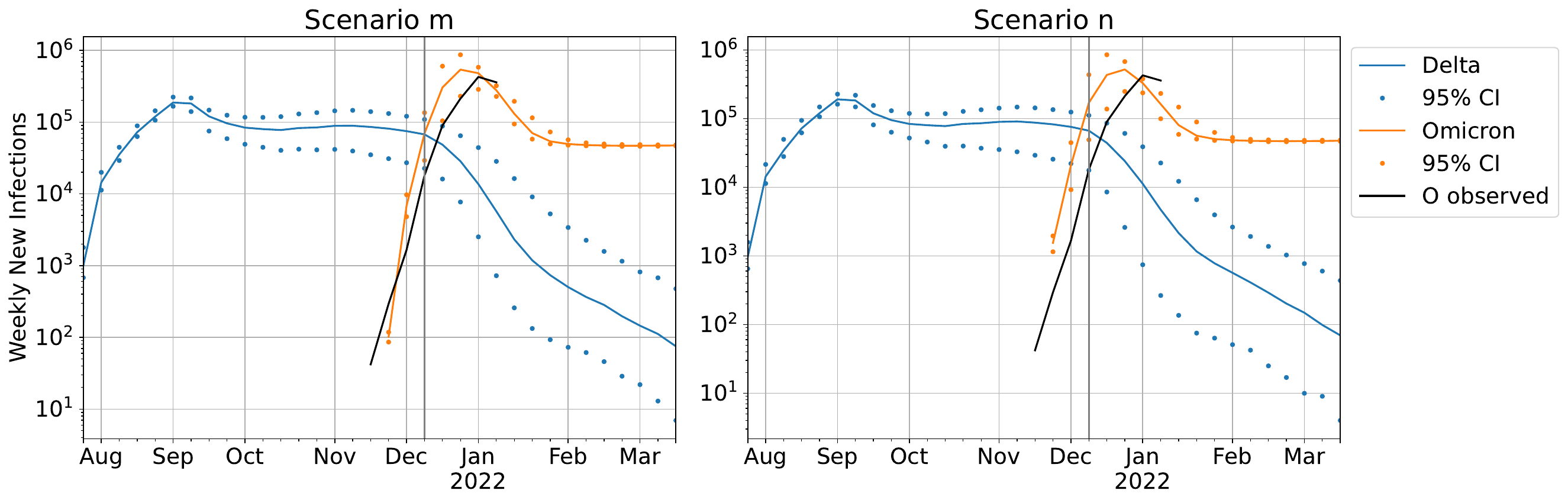}
  \includegraphics[width=0.49\textwidth]{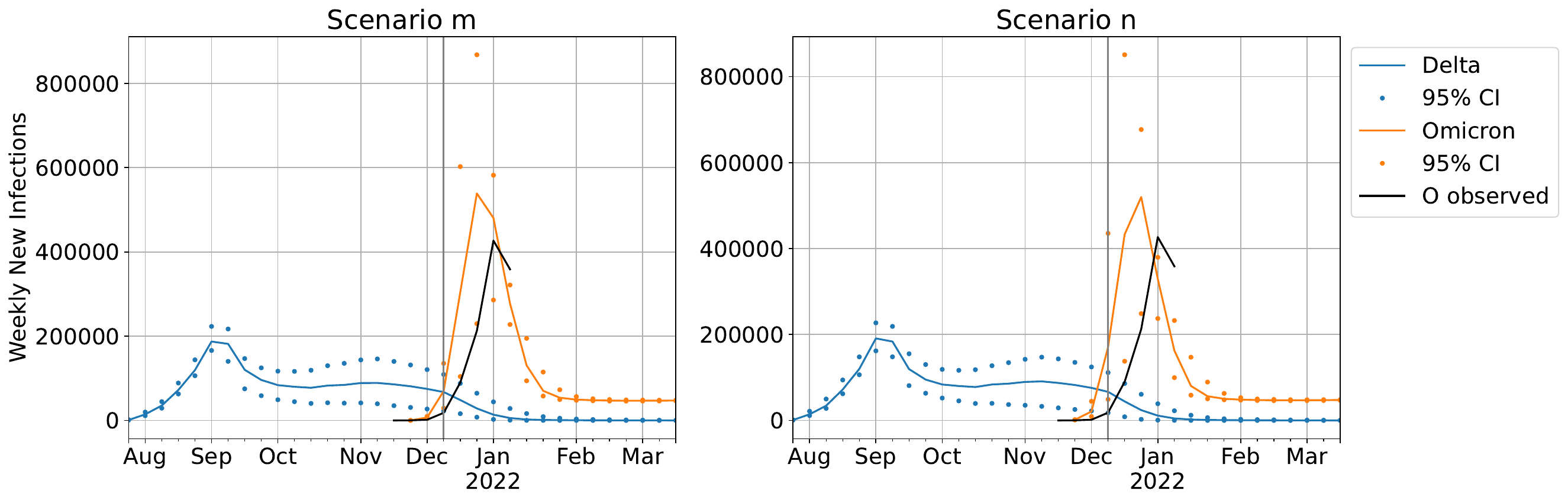}\\
  Explicit (knowledge post 11th Dec) distribution:\\
  \includegraphics[width=0.49\textwidth]{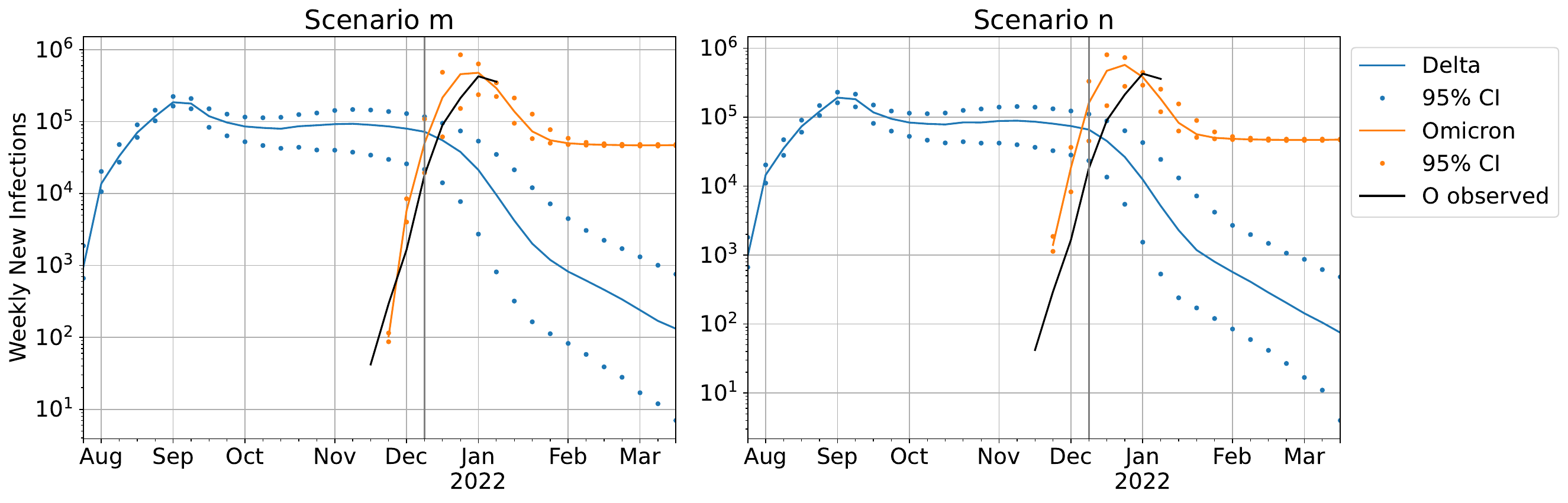}
  \includegraphics[width=0.49\textwidth]{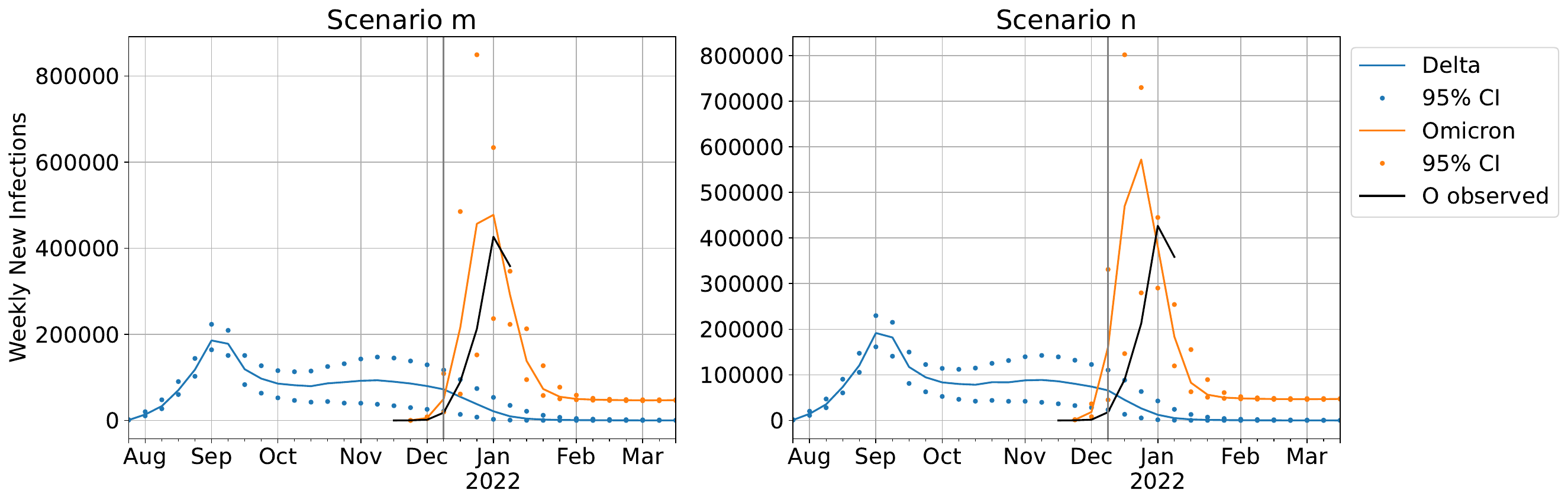}\\
  \caption{Adjusted transmission rate scenarios (log scale left, linear scale right) for each vaccine distribution scheme with lower vaccine escape (left of each pair) and higher vaccine escape (right of each pair). Dots represents bounds for 95\% of 200 simulations. Observed values prior to the vertical line (11th December) were the data available at the time of fitting, used to fit the growth rate; observed values subsequent to this date were added later for comparison.}
  \label{fig:all:scenario-grid-mn}
\end{figure}

We saw little difference between the different strategies for forward distribution of vaccines (55\% and 100\% uptake of boosters with respect to the uptake of 2nd doses).  Figure~\ref{fig:all:scenario-grid-mn} shows in detail the effect of the different strategies, including a model with the actual uptake post-11th December for comparison.

Based on the calculated transmission rate multiplier, we estimated Omicron to have a transmission rate advantage over Delta of 5.3$\times$ in the lower vaccine escape scenario, and an advantage of 5.1$\times$ in the higher vaccine escape scenario.

\begin{figure}
  \centering
  \includegraphics[width=\textwidth]{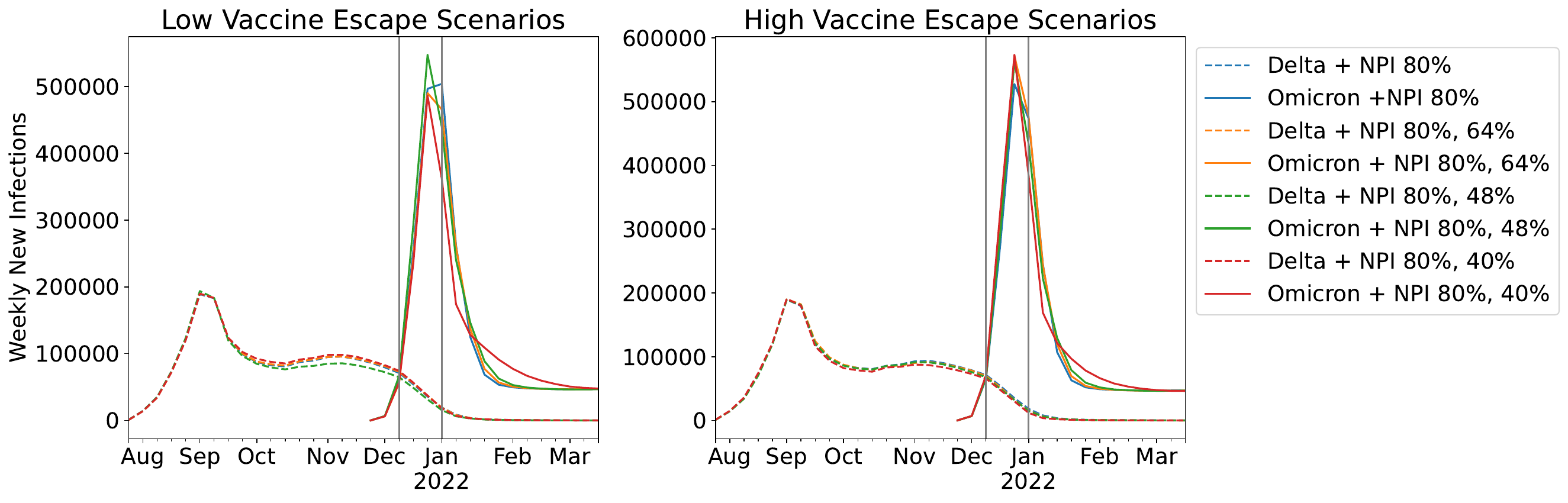}
  \caption{Adjusted transmission rate scenarios, lower vaccine escape (left), higher vaccine escape (right), with a range of post-Christmas NPI levels. Figure~\ref{fig:55:scenario-grid-mt} details the scenarios with confidence intervals.}
  \label{fig:55:scenario-comp-mt}
\end{figure}

\begin{figure}[h]
  \centering
  Linear scale:\\ \includegraphics[width=\textwidth]{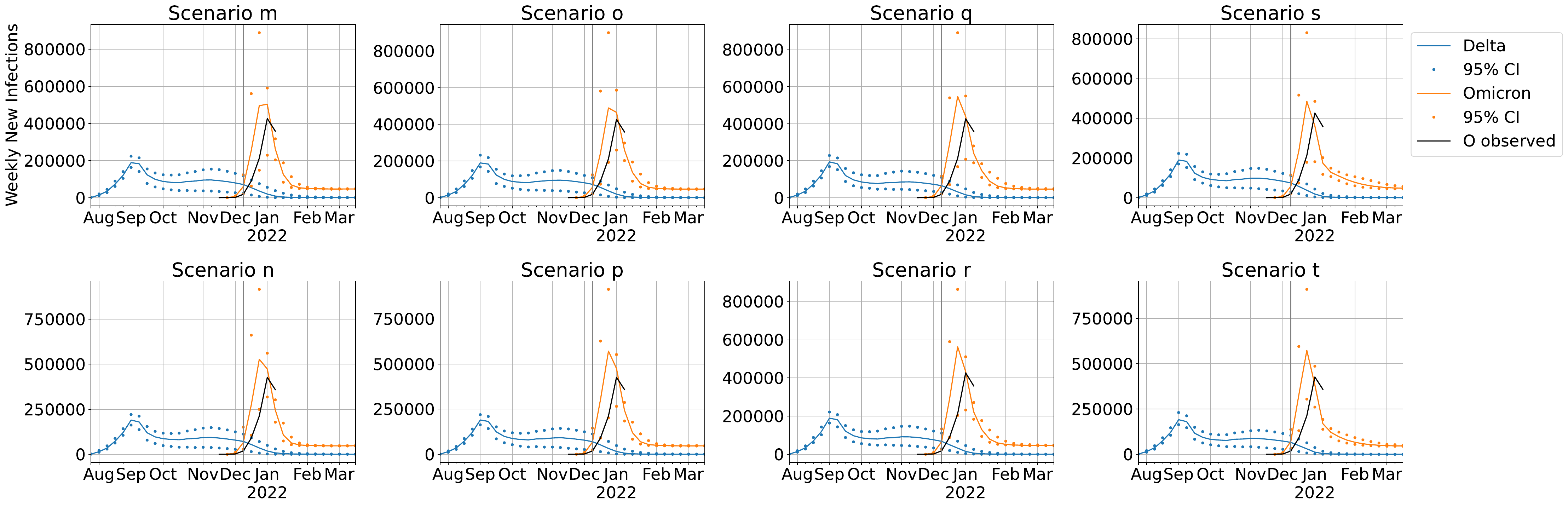}\\
  Log scale:\\ \includegraphics[width=\textwidth]{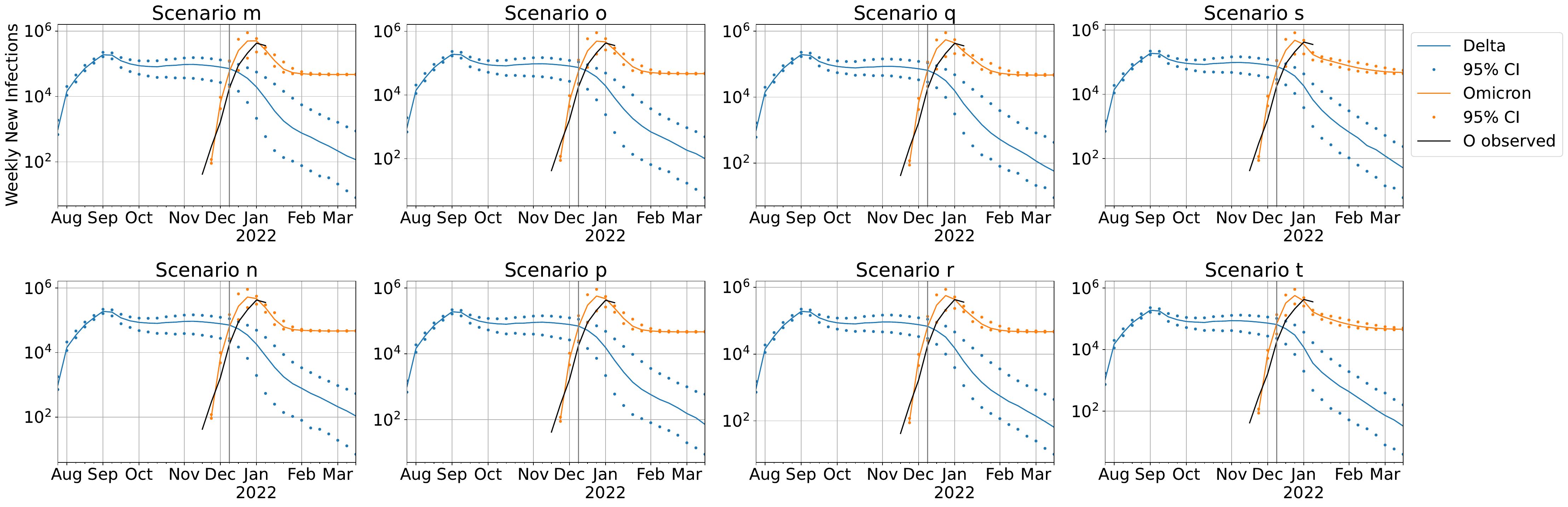}
  \caption{NPI adjusted transmission rate scenarios with 55\% uptake vaccine distribution scheme.}
  \label{fig:55:scenario-grid-mt}
\end{figure}

Not even the severest restrictions considered (a reduction of transmission rate to 40\% of the rate in mid November) resulted in a substantial reduction in cases, implying that restrictions alone were not likely to be sufficient to reduce the reproduction number below one at that time. Figure~\ref{fig:55:scenario-comp-mt} shows the range of post-Christmas NPI scenarios (central estimates). Figure~\ref{fig:55:scenario-grid-mt} details the NPI scenarios with confidence intervals. 

\subsection{Hospital occupancy estimates}
\begin{figure}
   \centering
   \includegraphics[width=1\textwidth]{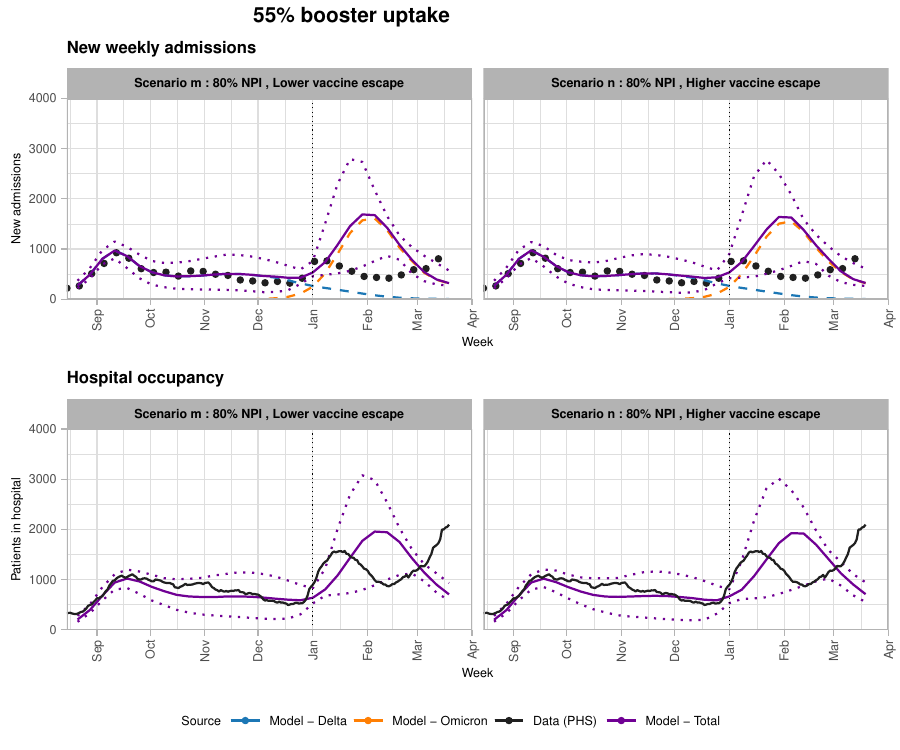}
   \caption{Estimated hospital admissions and occupancy from the infections trajectories in scenarios m and n, assuming the hospitalisation admission rates and occupancy for Omicron are identical to that of Delta.}
   \label{fig:hospital_occupancy_m_n}
\end{figure}

We estimate hospital admissions and hospital occupancy for Scenarios m and n in Figure~\ref{fig:hospital_occupancy_m_n}. We emphasise that we generated these trajectories on the assumption that the hospitalisation behaviour (admission rate, length of stay) for Omicron were identical to that of Delta between 1 May 2021 and 1 December 2021 (Table~\ref{tab:hosprates}); it later became clear that the case-to-hospitalisation rate of Omicron was in fact significantly lower than that of Delta~\cite{SHEIKH2022959}. The overall (purple) curves are combined admissions and occupancy from Omicron and Delta. We again include data of observed admissions (with the discrepancy between data and model confirming the lower severity) and occupancy past 11th December, not known at the time of original modelling.

\section{Discussion}
In this study we have adapted an existing model, that was used to provide the Scottish Government with medium term projections of cases of COVID-19 across Scotland over much of the ``emergency'' period of COVID-19 restrictions. While many of the results were similar to those that have been described by more parsimonious models, the spatially detailed individual-based model described here does have some advantages. In particular, the local spatial heterogeneity in distribution of prior immunity (from Delta) and heterogeneity on vaccine distribution and uptake are relevant when attempting the estimate the effect of vaccine escape and transmission rate advantage, and (while not directly considered in this analysis) therefore allows for direct consideration of the impact of, for example, targeting improved vaccination uptake on the basis of local cases numbers plus deprivation. 

The model was adapted to account for the transmission of two strains of SARS-CoV-2 with different levels of vaccine immune escape, in order to mimic the properties of the emergent Omicron VOC as known at time of analysis. Because this was a period of sustained consistent exponential growth, with no obvious change in interventions, this allowed us to fit a single model assuming a single baseline transmission rate. In the scenario models we showed that, under plausible levels of immune escape, only transmission rate advantages of between 5.1$\times$ and 5.3$\times$ result in trajectories consistent with the observed numbers of cases with SGTF (assumed to be the Omicron variant). These estimates are consistent with other estimates at the time (at the upper ranges, though well within the published CIs)~\cite{Ito2022,Viana2022}. While further combinations of transmission rate advantage and vaccine escape may of course be plausible, in our view the scenarios we examined provided a suitable backdrop on which to further investigate the impact of interventions beyond those imposed in mid-December and to provide reasonable mid-range forward projections of case numbers.

We saw little difference between the different vaccine uptake scenarios in the model. This is most likely because the landscape of immunity already present at the time of the Omicron outbreak has a substantially greater impact on transmission than the relatively small number of doses administered in the short period after the outbreak. Hence we consider a reasonable best estimate at projected distribution is sufficient for modelling a short to medium-term projection of infections. This would most likely not be the case for longer-term projections. While there is some evidence that immune protection is at best, partial \cite{Lind2023}, it is difficult to translate the outcomes from the correctional institute setting of these data to our analysis of general circulation in Scotland. Further it is not likely to have a substantial impact over short and medium term projection periods though it does place important restrictions on the validity of longer term projections.

We also saw that the imposition of post-Christmas NPI restrictions that were likely to have little effect, firstly because the transmission rate advantage of the Omicron variant was so high, but also because post-Christmas the rapid outbreak would likely have already peaked meaning any restrictions at that time would likely have come too late.

One important assumption in the model was that vaccinated individuals who become infected, were as likely to experience severe disease as others. While it is known that vaccines do provide very good protection against severe infection with the Delta VOC, such data were not available at the time for the Omicron VOC. With its observed greater ability to evade vaccine induced protection, this was chosen as the more conservative option. We also assumed complete cross immunity between strains. It is known now that that this is not the case, but again there was little data available at the time to make an assessment on this. Finally, we assumed no change in virulence or outcomes. Whilst estimates from South Africa were available, there was also a very different immunity landscape and a much lower vaccine uptake in that region, so we felt that we could not compare well enough to make assumptions.


We observe that our final adjusted scenarios remained consistent with observed Omicron cases beyond the observations available at the time of fitting. Despite assumptions, we have shown that it was possible, with a model that had been developed over the course of the Delta variant period to make a rapid and reasonable estimate of the impact of a new variant with little knowledge about its detailed dynamics, an analysis aided by the tracking of immunity throughout the population during the Delta period at fine geographical scale, and thus facilitating the model's ability to simulate the finer detail of the spatial spread of the new variant.

A retrospective comparison with the observation data available after the analysis period shows that the model's projection of both the scale and timing of the Omicron outbreak and the likely utility, or lack thereof, of post-Christmas NPIs were consistent with the outcome data and therefore could be considered predictive. This indicates that for that time period, the landscape of immunity prior to a variant outbreak, if it has a significant transmission rate advantage, was much more important for modelling the dynamics of the initial outbreak than any shorter-term change in immunity or contact patterns after the outbreak has taken hold. We have shown that how a computationally intensive individually-based spatial simulation model was used under conditions of urgent, policy-relevant needs, and that this model provided robust advice to the Scottish government. Such an approach does come with some cost. The model complexity can make model alterations unwieldy, though in this case, this was mitigated by the availability of a dedicated software developer. It also is expensive in terms of compute time and resource, and this is exacerbated by the challenges of parameter inference with a large number of fitted parameters. 

Using this framework, further work could be done to improve the model's capacity to generate projections. A previous analysis~\cite{Wood2023} has identified the importance of finer-grained age demographics, household sizes, and student populations in determining the distribution of cases over the whole delta and omicron periods. Therefore the inclusion of these factors are likely to be important for the observed patterns that our model captures. Further work could be done to refine the simulation approach along these lines, however this lies beyond the scope of this paper.

While similar conclusions could be made with more parsimonious approaches, our spatially explicit model would have been well set up to examine more refined interventions such as improving vaccination uptake in deprived areas, should there have been evidence this was merited. By building upon a professionally developed code base and with the availability of appropriate high-granularity data, we were able to provide timely advice to the Scottish government, on timescales similar to other teams, providing support for the development and utility of similar approaches in the future.


\bibliographystyle{unsrt}
\bibliography{SCoVMod}


\clearpage
\beginsupplement
\section*{Supplementary Figures}



\begin{figure}[h]
  Within each OA $i$ the infection process is governed by the frequency dependent force of infection at time $t$:
    \begin{align*}
    \Lambda_i(t) = \Bigg[\Bigg. & \left(\beta _N+\beta _D\right)\left(\sum _{a'{\in}Y,E}\left(yI_{ia'}^M+I_{ia'}^S\right)\right)+\beta
                                     _N\left(yI_{ia}^M+I_{ia}^S\right)+\beta _DI_{ia}^S \\
                                   & +\beta _D\Bigg\{\Bigg.\Big\{\Big.\sum _j(1-\sum
                                     _{i'}x_{\mathit{i'j}})(yI_{i'a}^M)\Big\}\Big.\\
                                 &\hspace{4em} +\sum _j\Big\{\Big.x_{\mathit{ij}}\Big(\Big.\sum
                                     _k(1-x_{\mathit{jk}})(yI_{ja}^M)+I_{ja}^S+\sum
                                     _{a'{\in}Y,E}(yI_{ja'}^M+I_{ja'}^S)\Big)\Big.\Big\}\Big.\Bigg\}\Bigg. \Bigg]\Bigg. /N_i
  \end{align*}

  where:
  \begin{itemize}
  \item $\beta_N$ and $\beta_D$ are the nighttime and daytime transmission rates, 
  \item $a \in \{Y,A,E\}$ is the age class (Young, Adult, Elderly), 
  \item $y$ is the transmission rate modifier for mildly infected individuals, 
  \item $I_{ia}^M$ is the number of mildly infected at location $i$ in age class $a$, 
  \item $I_{ia}^S$ is the number of severely infected at location $i$ in age class $a$, 
  \item $x_{ij}$ is the proportion of individuals commuting between locations $i$ and $j$, 
  \item $N_i$ is the population of location $i$.
  \end{itemize}
  \caption{Equation: Force of infection for location $i$ at time $t$.}
  \label{fig:force}
\end{figure}


\begin{table}[h]
  \scriptsize{
  \begin{tabular}{|p{0.3\textwidth}|l|l|l|l|l|p{0.1\textwidth}|}
    \hline
    \textbf{Parameter}                                                             & \textbf{Transition} & \textbf{Symbol}     & \textbf{Age}     & \textbf{Value} & \textbf{Prior}      & \textbf{References}  \\
    \hline\hline
    Latency period                                                         & $E \rightarrow I^M$     & $1/\gamma$        & All     & fitted                             & U(1.67,28) & \cite{Arenas2020,He2020,Li2020,Bi2020,Zhang2020}          \\
    \hline
    Days from mild infectiousness to recovery                              & $I^M \rightarrow R$     & $1/\rho_M$       & All     & fitted                             & U(0.67,28) & \cite{DiDomenico2020,Zhang2020}          \\
    \hline
    Symptom onset time after infectiousness                                & $I^M \rightarrow I^S$    & $1/\gamma_M$       & All     & fitted                             & U(2,28)    & \cite{Arenas2020,He2020,Li2020,Bi2020,Zhang2020,Linton2020,Lauer2020,Sanche2020,Sanche2020a}          \\
    \hline
    Transmission rate for severe infectors (baseline, daytime)             & $S \rightarrow E$      & $\beta_d$         & All     & fitted                             & U(0,6)   &           \\
    \hline
    Transmission rate for severe infectors (baseline, nightime)            & $S \rightarrow E$      & $\beta_n$         & All     & fitted                             & U(0,6)   &           \\
    \hline
    Transmission rate multiplier for mild infectors                        & $S \rightarrow E$      & $y$          & All     & fitted                             & U(0,2.6)   &           \\
    \hline
    Severe symptom onset to hospitalization                                & $I^S \rightarrow H$     & $1/\eta$        & All     & 4                                  &           & \cite{Sanche2020,Sanche2020a,Wang2020,Liu2020,Han2020,Pung2020,Verity2020,Filipe2020}          \\
    \hline
    Severe symptom onset to recovery                                       & $I^S\rightarrow R$     & $1/\rho_S$       & Young   & 19                                 &           & \cite{Sanche2020}          \\
    for non-hospitalised                                                                      &            &            & Adults  & 20.7                               &            &            \\
                                                                                   &            &            & Elderly & 21.6                               &            &            \\
    \hline
    Days hospitalisation to death                                          & $H \rightarrow D$      & $1/\mu_H$       & Young   &                                   &           & eDRIS data \\
                                                                                   &            &            & Adults  & 6.97                               &            & eDRIS data \\
                                                                                   &            &            & Elderly & 6.62                               &            & eDRIS data \\
    \hline
    Proportion of hospitalised who                                         & $H \rightarrow R$      & $\rho_H/(\rho_H+\mu_H)$ & Young   & 1                                  &           & eDRIS data \\
    recover                                                                &            &            & Adults  & 0.96                               &            & eDRIS data \\
                                                                                   &            &            & Elderly & 0.84                               &            & eDRIS data \\
    \hline
    Symptoms onset to death                                                & $I^S \rightarrow D$     & $1/\mu_S$       & Adults  & 16                                 &           & \cite{Sanche2020a,Qin2020,Liu2020,Arentz2020,Filipe2020,Verity2020}         \\
    \hline
    Mortality rate multiplier (relative to average health index)           &            & $\mu_{mod}$       & All     & fitted                             & U(0,0.08)  &            \\
    \hline
    Number of seed infections (Delta variant)                                             &            & $N_s$         & N/A     & fitted                             & U(10, 20000) & \\
    \hline
    Number of seed recovered individuals (Delta variant)                                             &            & $N_R$         & N/A     & fitted                             & U(2m, 5m) & \\
    \hline
  \end{tabular}
  }
  \caption{Epidemiological parameters in SCoVMod, with priors and
    fixed values as appropriate. Where age is not indicated,
    parameters are assumed to be age independent. All times are
    measured in days.}
  \label{tab:params}
\end{table}

\begin{table}[h]
\begin{tabular}{lrrrr}
  \hline
Age range & Cases & \% & Admissions & Rate (\%) \\ 
  \hline
0-4 & 11371 & 2.40 & 495 & 4.40 \\ 
  5-9 & 38038 & 8.00 & 128 & 0.30 \\ 
  10-14 & 55824 & 11.70 & 142 & 0.30 \\ 
  15-19 & 49391 & 10.30 & 201 & 0.40 \\ 
  20-24 & 45189 & 9.40 & 299 & 0.70 \\ 
  25-29 & 38415 & 8.00 & 421 & 1.10 \\ 
  30-34 & 34108 & 7.10 & 565 & 1.70 \\ 
  35-39 & 33708 & 7.00 & 678 & 2.00 \\ 
  40-44 & 35242 & 7.40 & 655 & 1.90 \\ 
  45-49 & 29724 & 6.20 & 646 & 2.20 \\ 
  50-54 & 28719 & 6.00 & 791 & 2.80 \\ 
  55-59 & 25186 & 5.30 & 918 & 3.60 \\ 
  60-64 & 19220 & 4.00 & 951 & 4.90 \\ 
  65-69 & 11770 & 2.50 & 837 & 7.10 \\ 
  70-74 & 8968 & 1.90 & 1060 & 11.80 \\ 
  75+ & 13467 & 2.80 & 3335 & 24.80 \\ 
  \hline
  Total & 478340 & 100.00 & 12122 & 2.53 \\ 
   \hline
\end{tabular}
\caption{Hospitalisation rates over the period 1 May 2021 -- 1 December 2021, used to inform hospitalisation projections.}
\label{tab:hosprates}
\end{table}


\clearpage
\section*{Supplementary: Derivation of formula to estimate incidence from test data}
\label{incidence_estimate}
Testing data comes predominantly through the government's pillar two testing programme which required individuals in the population to make their own decision about when to take a test. Government guidance was to take a test if experiencing symptoms, living in the household of someone infected, or if recently in contact with someone infected\footnote{\url{https://www.gov.uk/guidance/nhs-test-and-trace-how-it-works}}. Studies show that the primary reasons for testing are the presence of symptoms, or a history of contact with an infected person.\footnote{
Supplisson et al. SARS-CoV-2 self-test uptake and factors associated with self-testing during Omicron BA.1 and BA.2 waves in France, January to May 2022. Euro Surveill. 2023;28(18):pii=2200781 \url{https://www.gov.uk/government/publications/lfd-tests-how-and-why-they-were-used-during-the-pandemic/covid-19-general-public-testing-behaviours}}

Therefore, for each infection we can expect multiple individuals to be tested locally. For this reason we expect the disproportionate levels of testing in a given area to be an indicator of higher levels of infection in that area. Other influencing factors, such as employment and education level, have not been included for simplicity. While these factors influence testing uptake at  an individual level,\footnote{Smith et al. Who is engaging with lateral flow testing for COVID-19 in the UK? The COVID-19 Rapid Survey of Adherence to Interventions and Responses (CORSAIR) study BMJ Open 2022;12:e058060} it is unclear whether the same effects would be present when aggregated across a region (unlike the disease incidence which is expected to exhibit strong spatial clustering).

The reason to include this mechanism is to moderate between two problematic methods: either using the number of positives as the main indicator of incidence, or taking the positivity rate (the proportion of people tested who give a positive result). Some regions can have a small number of cases yet a very large positivity rate if few people have been tested. In this case the estimated incidence will give contradictory outcomes depending on which method you apply. In our method the use of test numbers as a third indicator provides an estimate somewhere in between the two extremes.

We focus on one region of the nation at a given time interval. There are three pieces of information we will utilise for the region: its population, $n$; the number of tests conducted, $s$; and the number cases, $c$, defined as the individuals who tested positive. Additionally, we use $S$ and $C$ to denote the number of tests and cases, respectively, across the entire nation during the same time interval. Supposing the the national ascertainment rate is $a$, we estimate the number of infections, $I$, to be $I=C/a$.

Our aim is to find a probability distribution for the number of test-sensitive individuals at time $t$, defined as the number of individuals who would test positive on that day if they were tested. We calculate the probability distribution for the proportion of individuals in the region who are test-sensitive, $x$, as a function of these data, denoted $P(x|c,s,n)$. The mean of this distribution forms our estimate of prevalence in the region at the given time. $P(x|c,s,n)$ is composed of three probabilities: $P(x|P')$, the probability that a proportion $x$ are test-sensitive given the probability distribution $P'(x)$ at the previous time interval; $P(s|x;S,I)$, the probability that $s$ people were tested in the region expressed as a function of the number of test-sensitive people there, and the numbers of infected people and tests reported nationally; and $P(c|x;s,n)$, the probability that $c$ people tested positive, expressed as a function of the number who are test-sensitive, tested, and the total.

Using Bayes' formula We can express $P(x|c,s,n)$ as the product of three probabilities and a normalising constant
\begin{equation}
\label{three_parts}
    P(x|c,s,n)=P(c|x;s,n)\times P(s|x;S,I)\times P(x|P')\times\text{Const.}.
\end{equation}
Each term on the right hand side of this equation can be derived from assumptions about the relationship between incidence and test data, described below.

For $P(i|P')$, we make the assumption that the prevalence a the current time interval will be influenced by the prevalence in the previous time interval. We assume $P'(x)$ follows a gamma distribution $\Gamma(\alpha,\beta)$, explicitly $P'(x)=x^{\alpha-1}e^{-\beta x}\times\text{C}$ where $C$ is a normalising constant. This form of distribution is particularly convenient for reasons that will become clear later. 

Before utilising the test data, we have no knowledge of whether infections have increased or decreased. We therefore assume the mean $\mu=\alpha/\beta$ of the distribution is unchanged, and, since we accept that some change has occurred, that the standard deviation $\sigma=\alpha/\beta^{2}$ has increased by some amount. We choose to increase the standard deviation by a factor of $1/z$ where $z$ is chosen between $0$ and $1$ to control the amount that the previous time interval influences the current one. This is achieved by transforming the parameters $\alpha\rightarrow z\alpha$ and $\beta\rightarrow z\beta$, giving
\begin{equation}
    P(x|P') = x^{z\alpha-1}e^{-z\beta x}\times\text{Const.}
\end{equation}

The second main assumption is that the number of tests conducted in a given region reflects the prevalence of infected individuals. This follows from the non-random way in which tests are distributed; before requesting a test, most individuals have symptoms or another reason to believe that they have been exposed. To model this, we consider the probability, $q$, that any one of the $S$ tests conducted nationally is conducted on an individual in the region. We then assume that $q$ is proportional to the fraction of national infections that are to be found in the region, thus $q=nx/I$. The probability that exactly $s$ of the $S$ tests are conducted in the given region is  
\begin{equation}
    P(s|x) = \binom{S}{s}\left(\frac{nx}{I}\right)^{s}\left(1-\frac{nx}{I}\right)^{S-s}.
\end{equation}
The Poisson approximation to the binomial is valid here since $S$ is large and $nx/I$ is small, giving
\begin{equation}
    P( s | x )\approx x^{s}e^{-(nS/I)x}\times\text{Const.}
\end{equation}

The final assumption is that the number of individuals who test positive is a reflection of the number who are test-sensitive in the population. To do this we assume the probability that a test conducted in the region will be positive is proportional to the number of test-sensitive people there. The probability that $c$ of $s$ tests give a positive result is then
\begin{equation}
    P(c|x;s) = \binom{s}{c}x^{c}(1-x)^{s-c}
\end{equation}

The Poisson approximation to the binomial is valid since $s$ is large and $x$ is small, giving
\begin{equation}
    P( c | x ; s)\approx x^{c}e^{-sx} \times\text{Const.}
\end{equation}
Returning to Eq.~\eqref{three_parts} we can now write
\begin{equation}
\label{new_gamma}
P(x|c,s,n)=x^{c+s+z\alpha-1}e^{-(s+nS/I+z\beta)x}\times\text{Const.}
\end{equation}
where $\alpha$ and $\beta$ are the parameters for the equivalent gamma distribution at the previous time interval. Noticing that Eq.~\eqref{new_gamma} has the form of a gamma distribution, $\Gamma(c+s+z\alpha,T+nS/I+z\beta)$ we see from one time interval to the next the form of the distribution for $x$ does not change, but the parameters $\alpha$ and $\beta$ do.

Using $\alpha_{t}$ and $\beta_{t}$ to denote the Gamma distribution parameters for $x$ at time $t$, and similarly using the same subscript to denote the values of $I$,$s$ and $c$ to refer to their values at the time interval indicated, we have that $\alpha_{t}=c_{t}+s_{t}+z\alpha_{t-1}$ and $\beta_{t}=s_{t}+nS_{t}/I_{t}+z\beta_{t-1}$. Arbitrarily choosing $\alpha_{0}=\beta_{0}=1$ we can write $\alpha_{t}=\sum_{i=1}^{t}(s_{i}+c_{i})z^{t-i}$ and $\beta_{t}=\sum_{i=1}^{t}(s_{i}+nS_{i}/I_{i})z^{t-i}$. Since the mean of a Gamma distribution is $\alpha/\beta$ we have
\begin{equation}
    \mu_{t}=\frac{\displaystyle \sum_{i=1}^{t}(s_{i}+c_{i})z^{t-i}}{\displaystyle \sum_{i=1}^{t}(s_{i}+nS_{i}/I_{i})z^{t-i}}
\end{equation}
which is our estimate of the proportion of the population who are test-sensitive at time $t$.


\clearpage
For the purpose of open access, the author has applied a Creative Commons Attribution (CC~BY) licence to any Author Accepted Manuscript version arising from this submission.

\end{document}